\documentclass[twocolumn,amssymb,amsmath,floats,showpacs,pre]{revtex4-1}

\usepackage{amsmath,amssymb,bm}
\usepackage{graphicx}
\usepackage{xcolor}
\usepackage[toc,page]{appendix}
\usepackage{comment}
\usepackage[unicode=true]{hyperref}
\usepackage{enumerate}
\pdfoutput=1

\newcommand{\av}[1]{\left\langle {#1} \right\rangle}

\begin{document}

\title{Competition between vaccination and disease spreading}

\author{Mozhgan Khanjanianpak}
\affiliation{Physics Department, Institute for Advanced Studies in Basic Sciences, 45195-1159 Zanjan, Iran}

\author{Nahid Azimi-Tafreshi}
\affiliation{Physics Department, Institute for Advanced Studies in Basic Sciences, 45195-1159 Zanjan, Iran}

\author{Claudio Castellano}
\affiliation{Istituto dei Sistemi Complessi (ISC-CNR), via dei Taurini 19, I-00185 Roma, Italy}

\begin{abstract}
We study the interaction between epidemic spreading
and a vaccination process.
We assume  that, similar to the  disease spreading, the vaccination
process also occurs through direct contact, i.e., it follows the standard
susceptible-infected-susceptible  (SIS) dynamics.
The two competing processes are asymmetrically coupled as
vaccinated nodes can directly become infected at a reduced rate
with respect to susceptible ones.
We study analytically the model in the framework of mean-field theory
finding a rich phase diagram.
When vaccination provides little protection toward infection,
two continuous transitions separate a disease-free immunized
state from vaccinated-free epidemic state, with an intermediate
mixed state where susceptible, infected, and vaccinated individuals coexist.
As vaccine efficiency increases, a tricritical point leads to
a bistable regime, and  discontinuous phase transitions emerge.
Numerical simulations for homogeneous random networks agree very
well with analytical predictions.
\end{abstract}


\maketitle

\section{Introduction}

The study of disease spreading in well-mixed and networked populations
has attracted much interest in recent years~\cite{Newman, castellano}.
To understand disease dynamics, many
mathematical models of epidemic spreading have been developed.
A paradigmatic role is played by the Susceptible-Infected-Susceptible (SIS)
model, in which nodes are in one of two possible states:
susceptible (S) or infected (I).
Each susceptible node gets infected, with probability $\beta$ per unit time,
through any of its connections to infected neighbors.
At the same time, each infected node spontaneously recovers
at rate $\mu$, returning to the susceptible state S.
Above a critical value of the ratio $\beta/\mu$ (epidemic threshold)
an endemic state with a finite fraction of infected nodes is reached,
while below the threshold the infection dies out exponentially fast.

‌To prevent or reduce the spread of a disease, different strategies have been proposed~\cite{vaccine-review, Pastor, Inverse}.
A primary and effective way to control epidemics is vaccination~\cite{Wang}.
Random vaccination doesn't need any information about the structure
of the network; however, it costs a lot and is inefficient when a limited
amount of resources are available.
Instead targeted vaccination based on the identification of the most
important nodes is more effective~\cite{target}. However, targeted
vaccination  requires global information about the structure of the network,
which is often unavailable.
To overcome this problem, acquaintance vaccination was proposed, in
which a fraction of nodes is selected at random, and then their neighbors
are randomly vaccinated~\cite{Acquaintance}.

In real cases, a vaccine may have only a transient effect, i.e.,
vaccinated individuals may return to the susceptible state after a
while (temporary vaccine). Also, the vaccination may not be
completely effective so that it is possible that a
vaccinated individual gets infected, even though at a smaller
transmission rate (leaky vaccine)~\cite{Gardon}. Some mathematical models were
introduced to take into account the effect of leaky and/or temporary
vaccines~\cite{Zaleta,Peng,Peng2,Steinegger,Chen}.
For instance, in Ref.~\cite{Peng} a third compartment
(vaccinated individuals, V) has been added to the SIS model:
a susceptible node can spontaneously get vaccinated at a given rate, and each
vaccinated individual can return to the S state with a susceptibility rate.
Furthermore, the authors considered a leaky vaccine such that
a vaccinated node can be infected at a reduced rate.
They studied the influence of imperfect vaccination on the threshold
and the reduction of epidemic prevalence in different networks.

In recent years the interaction between spreading processes,
in the case of both cooperation and competition among diseases,
has received much attention~\cite{Wang2, Karrer, Newman2, Cai, Cui, Min, Funk, Ruan,
Sahneh, Granell, Sanz, Azimi, Hang, Brodka}.
A part of these studies is concerned with the dynamical interplay
between a pair of diseases, spreading through the same network,
and investigates how one disease can promote or inhibit the spreading of the
other~\cite{Karrer, Newman2, Cai, Cui, Min}.
It is also possible that a disease competes with a preventing process,
such as the propagation of vaccination or the spreading of awareness
about the disease~\cite{Peng, Peng2, Funk}.
In particular in Ref.~\cite{Peng} the authors have studied the
competition between the propagation of a virus and the immunization
in an imperfect vaccination process. They ‌analyzed the possible
effects of vaccination on disease spreading occurring on various networks.
The interaction of multiple spreaders on multilayer networks, where each
spreader propagates on one layer, is more
complex~\cite{Sahneh, Granell, Sanz, Azimi, Hang, Brodka}.
On multilayer networks, coupling of spreading processes through interlayer
connections makes the transition point and the nature of the transitions
different.

In this paper we study the competition of disease spreading with
vaccination.  Similar to the model considered in Ref.~\cite{Peng},
we add a leaky and temporary vaccinated state to the SIS model.
While Ref.~\cite{Peng} assumes that susceptible individuals
can get spontaneously vaccinated at a given rate, in our model
we consider this transition as a contact process, i.e., susceptible
individuals may be convinced
to get vaccinated only if in contact with vaccinated neighbors.
In other words, we consider a three-state model and assume
that both the disease  and the vaccination propagate according to
the SIS dynamics.  In addition, we consider the vaccine to be imperfect,
so that  vaccinated individuals can get infected when in contact with
infected neighbors.
This possibility provides an additional coupling between the two
competing spreading processes.
Beyond the interpretation in terms of infection and vaccination,
our model can be seen as a generic model for two competing, mutually
exclusive, spreading processes, in the presence of a tunable dynamical
asymmetry~\cite{Yang2016,Wu2011,Wu2013,Ahn2006}.
To analyze the model behavior, we write dynamical mean-field
equations and solve them at
stationarity, deriving the rich phase diagram of the model.
As a function of model parameters we predict the existence of both continuous
and discontinuous transitions, separated by a tricritical point.
Below the tricritical point, a mixed state with coexistence of
susceptible, infected and, vaccinated individuals interpolates
between a state where the infection dies out and a state where
vaccination disappears. Interestingly, the mixed state turns out to exist
only in the presence of an asymmetry between the infected and
vaccinated state, i.e., only if a direct transition from vaccinated
to infected is possible.
Above the tricritical point the intermediate mixed state is
replaced by a bistability region, where the stationary state
depends on the initial condition.
We test these analytical results by performing numerical simulations on
random homogeneous networks, and we find a very good agreement.

The paper is organized as follows. In the next section, we define our
model, and, within the framework of mean-field theory, we find the fixed
points of the dynamics and analyze their stability.
We obtain the bifurcation diagrams for the model and show that bistability
emerges above a tricritical point.
In Sec. III we apply our results to homogeneous random networks
and compare them with numerical simulations.
In Sec. IV we present some concluding remarks and perspectives.

\section{The model and its mean-field analysis}

\begin{figure}[t]
	\includegraphics[width=0.6\columnwidth]{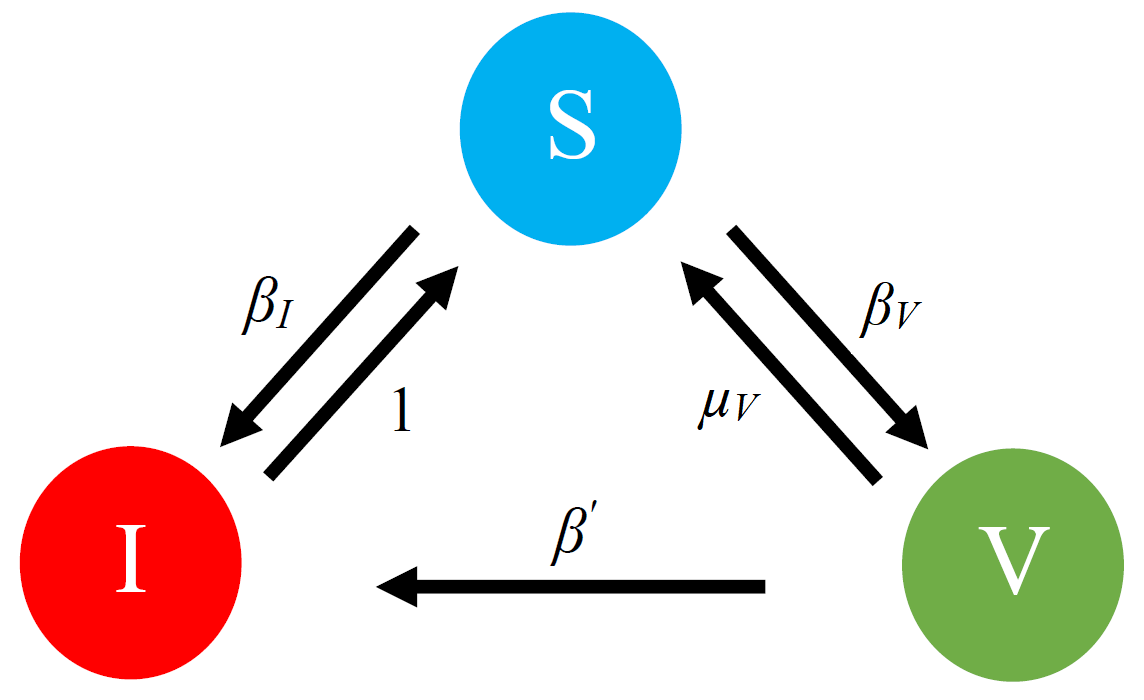}
	\centering
	\caption{Schematic representation of dynamics of the
          model. The rates indicated are for
            the transition of a single individual. Transitions
            from I to S and V to S are spontaneous. The other
            three possible transitions occur through direct contacts
            between the individual changing state and a neighbor
            inducing the transition.}
	\label{f1}
\end{figure}

Let us consider a susceptible-infected-vaccinated (SIV) model
for disease spreading, a SIS model modified to include a
vaccinated state.
Each node can be in one of three states: susceptible (S), infected (I), and
vaccinated (V).
The spreading of the infection and of the vaccination both take place
according to the SIS dynamics:
A susceptible node can acquire the infection from each of its infected
neighbors, with a probability per unit time $\beta_{I}$ per neighbor.
An infected node spontaneously recovers and becomes susceptible again
with a rate $\mu_I$, that we set equal to 1 with no loss of generality.
A susceptible node can also become vaccinated with a rate $\beta_{V}$,
through contact with a vaccinated neighbor.
A vaccinated node can lose its immunity and become
susceptible again with the rate $\mu_V$.
The processes just described are symmetric under the change
$I \leftrightarrow V$. This symmetry is broken by the possibility
that a vaccinated node gets directly infected by a neighbor.
This transition occurs at a reduced rate $\beta' < \beta_I$.
We write $\beta '=\beta_I (1-e)$ where $0 \le e \le 1$
is the efficiency of the vaccination.
The opposite transition, from I to V, is forbidden.
The dynamics of the model is summarized as follows (see also Fig.~\ref{f1}): 

\[\begin{array}{l}
S + V\xrightarrow{\beta_V} V + V,\\
S + I\xrightarrow{\beta_I} I + I,\\
V + I\xrightarrow{\beta'} I + I,\\
I\xrightarrow{1} S,\\
V\xrightarrow{\mu_V} S.
\end{array}\]

Let us denote the fractions of susceptible, infected, and vaccinated
nodes with $x$, $y$, and $z$, respectively. Since the number of nodes,
$N$, is constant, there is a conservation rule as $x+y+z=1$, and so
$\dot x + \dot y + \dot z = 0$.

According to the dynamics of the
model, we can write down the following mean-field (MF) equations,
which correspond to assuming that, at each time, each node interacts
with a single other node selected randomly:
\begin{eqnarray}
\label{eq1}
\dot x &=&  - \beta_V x(1 - x - y) - \beta_I xy + \mu_V (1-x-y)+y,~~~~ \\
\dot y &=& \beta'y(1 - x - y) + \beta_I xy - y.~~~~
\label{eq2}
\!\!\!\!\!
\end{eqnarray}
The first term on the right-hand side of Eq.~(\ref{eq1}) corresponds to the
vaccination process ‌and accounts for the conversion of susceptible nodes into
vaccinated ones with rate $\beta_V$, while the second term describes
the infection process with the conversion of susceptible nodes into infected
ones at rate $\beta_I$. The third and fourth terms account for the
recovery from the vaccinated and infected state back to the
susceptible state, occurring with rates $\mu_V$ and $1$, respectively.
Similarly, the first term on the right hand side of
Eq.~(\ref{eq2}) corresponds to the infection of vaccinated nodes
(rate $\beta'$).
The second and third terms of Eq.~(\ref{eq2}) correspond to
the conversion of susceptible nodes into infected and $vice$ $versa$
according to the rules of the standard SIS dynamics.

To analyze stationary solutions of these equations, we determine the
fixed points of the system.
Imposing $\dot x= \dot y = 0$ leads to the following fixed points:
\begin{eqnarray}
(1) \rightarrow&~~(x_1^*,y_1^*,z_1^*) = &(1,0,0),\\
(2) \rightarrow&~~(x_2^*,y_2^*,z_2^*) = & \left(\frac{\mu_V}{\beta_V },0,1 - \frac{\mu_V}{\beta_V }\right),~~~ \\
(3) \rightarrow&~~(x_3^*,y_3^*,z_3^*) = & \left(\frac{1}{\beta_I },1- \frac{1}{\beta_I},0\right),~~~\\
(4) \rightarrow&~~(x_4^*,y_4^*,z_4^*) = & \bigg(\frac{\mu_V + \beta' - 1}{\beta' - \beta_I  + \beta_V },~~~\nonumber\\
&~~~~~& \frac{\mu_V(\beta_I  - \beta') + \beta_V \left( {\beta' - 1} \right)}{\beta'\left( {\beta' - \beta_I  + \beta_V } \right)},~~~\nonumber\\
&~~~~~& \frac{\beta'(1-\beta_I)+\beta_V-\mu_V\beta_I}{\beta'(\beta'-\beta_I+\beta_V)} \bigg).
\end{eqnarray}
The trivial fixed point 1 indicates the state in which all nodes are
susceptible, i.e., the absorbing state. Fixed point 2 corresponds to
a state in which there are no infected nodes, while there is a coexistence
of susceptible and vaccinated ones (``disease-free immunized'' state).
Fixed point 3 is perfectly analogous to fixed point 2 but now the
coexistence is between susceptible and infected nodes: it is the usual
active state of SIS dynamics  (``vaccinated-free epidemic'' state).
Finally fixed point 4 corresponds to a state in which the fraction of
susceptible, infected, and vaccinated nodes are all different from zero
(``mixed'' state).
The relevance of these fixed points for the SIV dynamics depends on
their stability and whether their coordinates are physical, i.e.,
within the range between 0 and 1. By stability we
intend that the stationary solutions must be deterministically stable
in the limit of infinite size.
The Jacobian matrix associated to the MF equations~(\ref{eq1}) and (\ref{eq2}) is

\begin{equation}
\resizebox{.98\hsize}{!}{$\begin{bmatrix}{2\beta_V x + y(\beta_V  - \beta_I ) -\mu_V - \beta_V } & {1 + x(\beta_V  - \beta_I )-\mu_V} \\{y(\beta_I  - \beta')} & {x(\beta_I  - \beta') - 2\beta'y + \beta' - 1} \end{bmatrix}.$}
\end{equation}

The analysis is made easier by first distinguishing between
the cases $\beta_V/\mu_V < 1$ and $\beta_V/\mu_V > 1$.
They correspond, respectively, to the inactive and the active phase of
the SIS dynamics for the vaccination process (alone) in mean-field.
In other words for $\beta_V/\mu_V < 1$ the vaccination rate is insufficient
to sustain the presence of a finite fraction of vaccinated
individuals in the system.
Even neglecting the possibility of V $\to$ I transitions,
the density of vaccinated nodes decreases and tends to zero spontaneously.
It is then reasonable to expect that the stationary state of the overall
system will be exactly the same of a normal SIS process for disease spreading.
In the case $\beta_V/\mu_V > 1$ instead, the vaccination process in isolation
would lead to a finite prevalence of vaccinated nodes. It is then interesting
(and nontrivial) to investigate how this interplays with the disease
spreading process.

\begin{figure}[h!]
	\includegraphics[width=0.6\columnwidth]{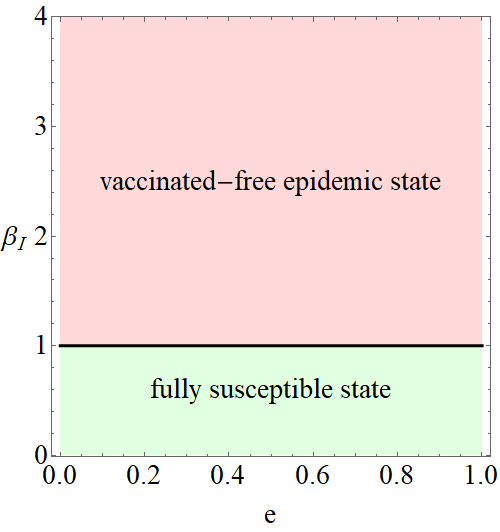}
	\centering
	\caption{Phase diagram of the model for $\beta_V/\mu_V<1$ as a
          function of $\beta_I$ and $e$. The solid line shows the
          continuous transition at $\beta_I=1$.}
	\label{f11}
\end{figure}

\subsection{The case $\frac{\beta_V}{\mu_V} < 1$}
\label{subsec1}

In this case, the fixed point 2 is not physical (as $x_2^*>1$),
so only three fixed points are relevant.

\subsubsection{Stability of fixed point 1}

The Jacobian matrix of the system for the first fixed point has
eigenvalues:
\begin{equation}
\left\{ \begin{array}{l}
{\lambda_1} = \beta_V -\mu_V, \\
{\lambda_2} = \beta_I-1.
\end{array} \right.
\label{eq3}
\end{equation}

In this regime, $\lambda_1$ is negative. Hence, in order for the fixed
point 1 to be stable, the infection rate $\beta_I$ must be smaller
than 1.

\subsubsection{Stability of fixed point 3}

The Jacobian matrix evaluated at the third fixed point
has the following eigenvalues:
\begin{equation}
\left\{ \begin{array}{l}
{\lambda _1} = 1 - \beta_I, \\
{\lambda _2} = \frac{{\beta_V  + \beta' - \mu_V\beta_I  - \beta_I \beta'}}{\beta_I }.
\end{array} \right.
\label{eq4}
\end{equation}
The first eigenvalue is negative if $\beta_I>1$.
The condition for the second to be negative is
\begin{equation}
\beta_I>\frac{\beta_V+\beta'}{\mu_V+\beta'} =
1 + \frac{\beta_V-\mu_V}{\beta'+\mu_V}.
\end{equation}
Since $\beta_V<\mu_V$,
for $\beta_I>1 $ both eigenvalues are negative and the
fixed point is stable.

\subsubsection{Stability of fixed point 4}
\label{subsec41}

The eigenvalues of the Jacobian matrix for the fourth fixed point are:

\begin{equation}
\left\{ \begin{array}{l}
{\lambda _{1}} = \frac{{(\mu_V\beta_I  - \beta_V)  - \sqrt {\frac{X}{{ - F}}} }}{{2\beta'}}, \\
{\lambda _{2}} = \frac{{(\mu_V\beta_I  - \beta_V)  + \sqrt {\frac{X}{{ - F}}} }}{{2\beta '}}.
\end{array} \right.
\end{equation}
where,
\begin{equation}
   \begin{aligned}
    &X = 4A\beta'^3+4B\beta'^2+5C\beta'+D, \\
    &A = (\mu_V  - \beta_V )(\beta_I  - 1),\\
    &B = (\mu_V\beta_I  - \beta_V )((1 - \beta_I)  + (\mu_V - \beta_V) ),\\
    &C =  - {(\mu_V\beta_I  - \beta_V )^2},\\
    &D = (\mu_V\beta_I -\beta_V)^2(\beta_I-\beta_V),\\
    &F = \beta_V  + \beta' - \beta_I.
    \label{eq6}
    \end{aligned}
\end{equation}

In order for the real part of both eigenvalues to be negative,
a necessary condition is ${(\mu_V\beta_I - \beta_V) } <0$
(independent of the value of $\sqrt{\frac{X}{-F}}$).
This condition means that $\beta_I<\frac{\beta_V}{\mu_V}$.
If this condition is satisfied, the real part of $\lambda_1$ is
necessarily negative. The real part of $\lambda_2$
is negative as well if $\frac{X}{{-F}} < (\mu_V\beta_I -\beta_V)^2$.
Whether this condition is fulfilled it depends on the sign of $F$.
For $F>0$, the condition is
satisfied if $ X>-F(\mu_V\beta_I - \beta_V)^2 $, which corresponds to
an inequality of the following general form:
\begin{equation}
A{{\beta'}^2} + B\beta' + C > 0.
\label{eq7}
\end{equation}
Since $\beta_V < \mu_V$, all coefficients $A$, $B$, and
$C$ are negative [Eqs.~(\ref{eq6})]. Hence, the inequality~(\ref{eq7})
is not satisfied for positive values of $\beta'$.  Therefore, the
fourth fixed point is never stable if $F>0$.
Instead for $F<0$, the sign of Eq.~(\ref{eq7}) is reversed, and the
inequality is always satisfied.
From the conditions $\beta_I<\frac{\beta_V}{\mu_V}$ and $F<0$
we obtain that the fourth fixed point is stable in the interval
$ \frac{\beta_V}{e} < \beta_I < \frac{\beta_V}{\mu_V} $. In this
interval we must check also that $x_4^*$, $y_4^*$, and $z_4^*$ belong to the
interval $(0,1)$. It can be proved that for all values of $e$,
$x_4^* $, and $y_4^*$ are never simultaneously physical (see Appendix A).
In summary, in the case $\beta_V < \mu_V$, the
fourth fixed point is never stable and physical at the same time.

We can conclude that, if $\beta_I<1$, the fixed point 1 (fully susceptible state)
is stable, while if $\beta_I>1$, the fixed point 3 (vaccinated-free epidemic
state)
is stable.
Therefore, as expected, in this regime the phase diagram is the same of the
standard SIS model for the spreading of a single disease (Fig.~\ref{f11}). The presence of
vaccinated individuals has effects only in the transient time before the
stationary state.

\begin{figure*}[t]
\begin{center}
\scalebox{0.6}{\includegraphics[angle=0]{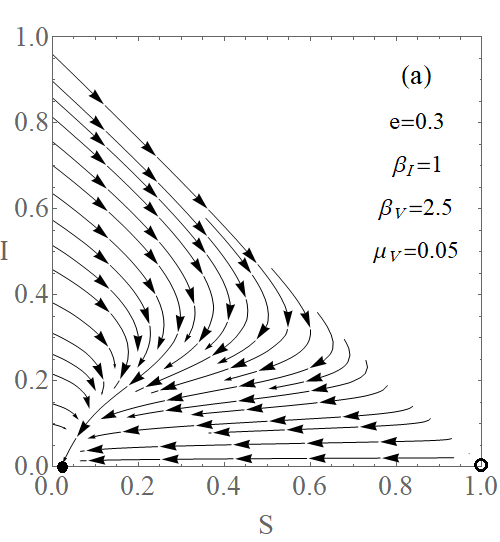}}~~~~
\scalebox{0.6}{\includegraphics[angle=0]{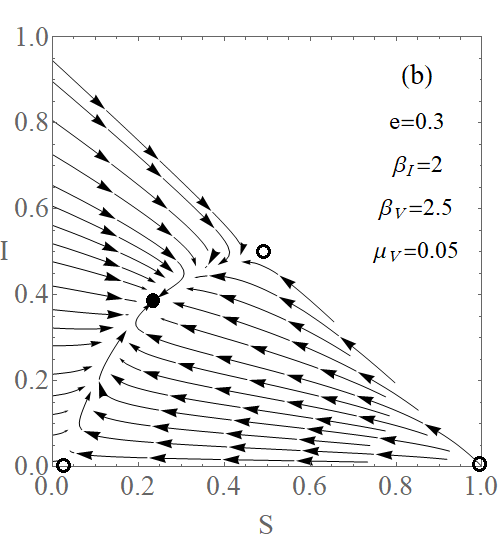}}~~~~
\scalebox{0.6}{\includegraphics[angle=0]{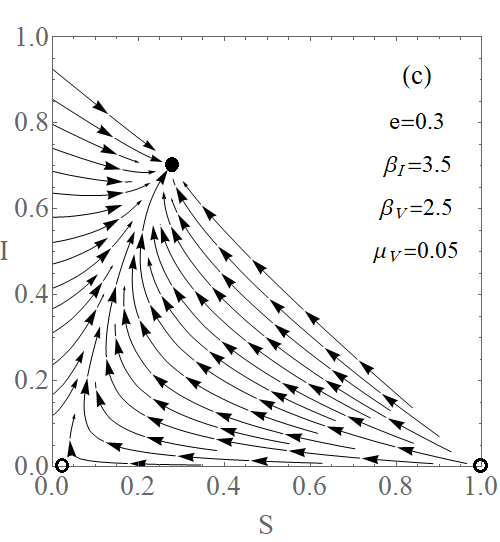}}~~~~
\end{center}
\caption{Phase portraits for $e<e_c$ and $(a)$ $\beta_I<\beta
  _V^*$, $(b)$ $\beta _V^*<\beta_I<\beta _E^*$, and $(c)$
  $\beta_I>\beta _E^*$. For the chosen values of $\beta_V$, $\mu_V$,
  and $e$, we find $\beta _V^*\simeq 1.416$, $\beta _E^*\simeq 2.410$,
  and $e_c\simeq 0.724$. Solid black dots represent stable fixed points, and open circles show saddle or unstable fixed points.  }
\label{f2}
\end{figure*}
\begin{figure*}[t]
\begin{center}
\scalebox{0.6}{\includegraphics[angle=0]{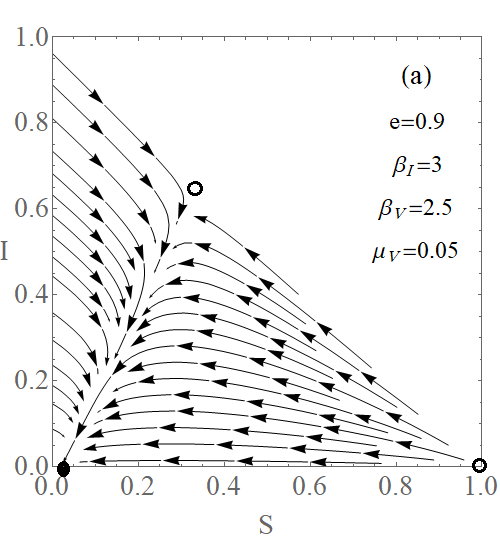}}~~~~
\scalebox{0.6}{\includegraphics[angle=0]{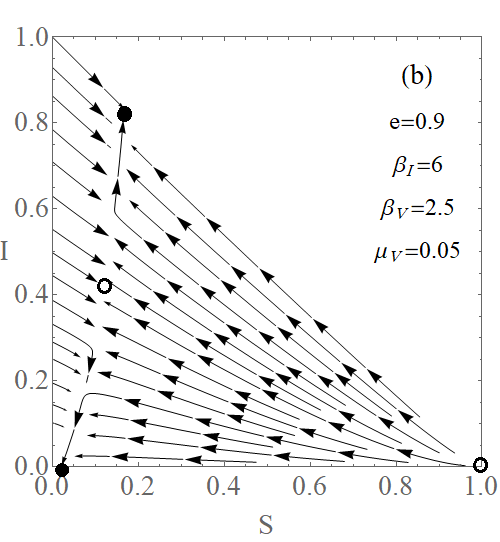}}~~~~
\scalebox{0.6}{\includegraphics[angle=0]{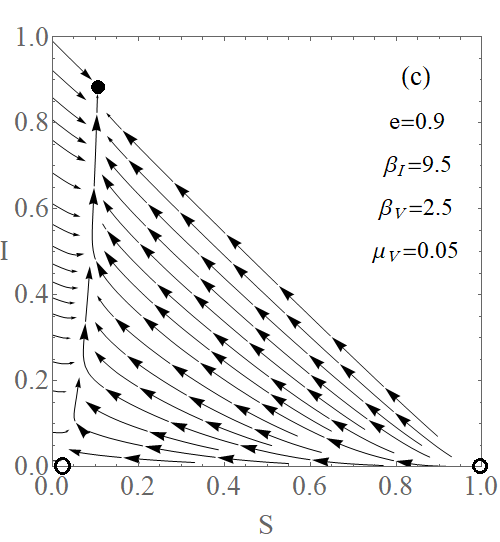}}
\end{center}
\caption{Phase portraits for $e>e_c$, $(a)$ $\beta_I<\beta _E^*$,
  $(b)$ $\beta _E^*<\beta_I<\beta _V^*$, and $(c)$ $\beta_I>\beta
  _V^*$. For the chosen values of $\beta_V$, $\mu_V$, and $e$, we find
  $\beta _V^*\simeq 8.474$, $\beta _E^*\simeq 5.256$, and $e_c\simeq
  0.724$. Solid black dots represent stable fixed points and open circles show saddle or unstable fixed points.}
\label{f3}
\end{figure*}

\subsection{The case $\frac{\beta_V}{\mu_V} > 1$}
\label{subsec2}

\subsubsection{Stability of fixed point 1}

The eigenvalues of the Jacobian matrix for this fixed point are given
by Eq.~(\ref{eq3}).  In this regime $\lambda _1$ is always positive,
hence this fixed point, which is a saddle node for
$\beta_I<1$ ($\lambda_2<0$), is never stable.

\subsubsection{Stability of fixed point 2}

The Jacobian matrix of the system for the second fixed point
has eigenvalues:
\begin{equation}
\left\{ {\begin{array}{*{20}{l}}
 	{{\lambda _1} = \mu_V - \beta_V },\\
 	{{\lambda _2} = \frac{{ - \beta_V  - \mu_V\beta ' + \mu_V\beta_I  + \beta_V \beta '}}{\beta_V }}
 	\end{array},} \right.
\end{equation}
$\lambda_1$ is always negative. Hence, the fixed point is
stable if $\lambda_2 <0$, which requires:
\begin{equation}
\beta_I  < \frac{\beta_V}{\beta_V(1-e)+e\mu_V} \equiv \beta_V^*.
\label{1con}
\end{equation}

\subsubsection{Stability of fixed point 3}

The eigenvalues of the Jacobian matrix evaluated for the third fixed point
are given by Eq.~(\ref{eq4}). This fixed point is physical for $\beta_I>1$,
a condition that guarantees that $\lambda _1<0$.
For having $\lambda_2<0$, the inequality
$\beta_I > \frac{{\beta_V + \beta '}}{{\mu_V + \beta '}}$ must hold.
By substituting $\beta '=\beta_I (1-e)$, we obtain the following inequality
for $\beta_I$:
\begin{equation}
f(\beta_I)\equiv {\beta_I ^2}(1 - e) + \beta_I (\mu_V + e - 1) - \beta_V  > 0.
\label{bequ}
\end{equation}
This inequality is satisfied for $\beta_I>\beta_E^*$, where
$\beta_E^*$ is the positive root of $f(\beta_I)=0$:
\begin{equation}
\beta_E^*\equiv\frac{(1-\mu_V-e) + \sqrt{(\mu_V+e-1)^2 + 4\beta_V(1-e)}} {2(1-e)}.
\label{2con}
\end{equation}
Therefore, the third fixed point is stable for $\beta_I>\beta_E^*$.

\begin{figure*}[t]
\begin{center}
\scalebox{0.3}{\includegraphics[angle=0]{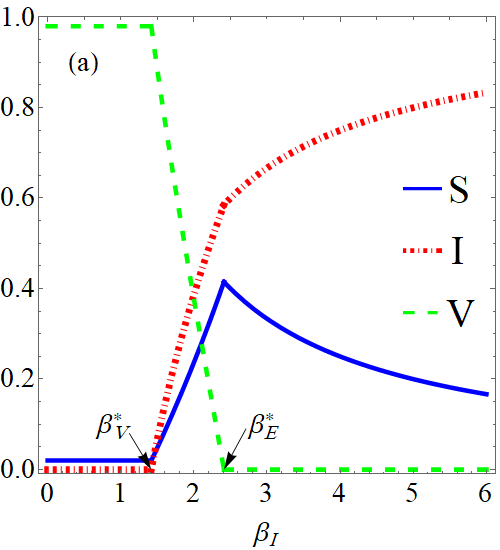}}~~~~
\scalebox{0.3}{\includegraphics[angle=0]{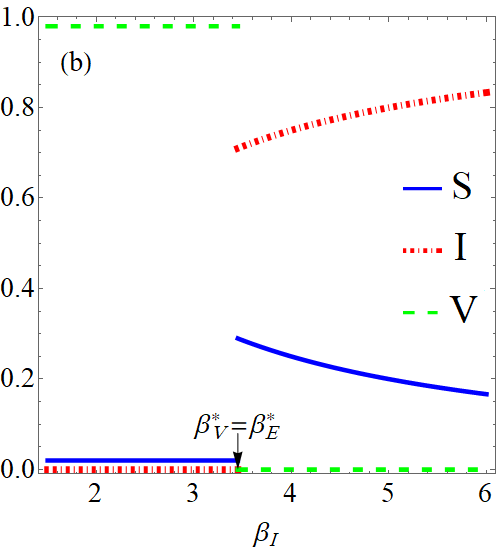}}~~~~
\scalebox{0.3}{\includegraphics[angle=0]{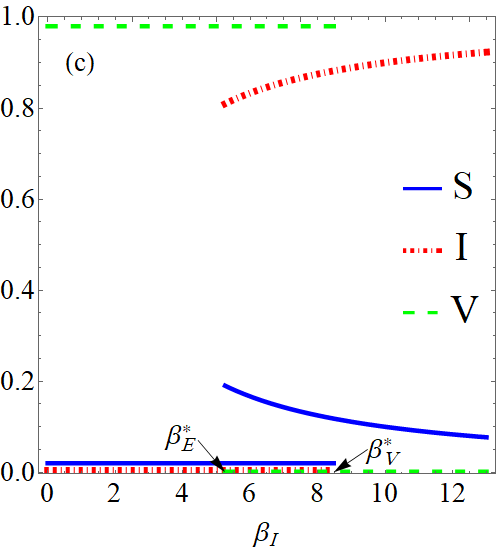}}~~~~
\end{center}
\caption{Stationary values of the
  fraction of susceptible, infected, and vaccinated nodes as a function
  of $\beta_I$ for $\beta_V=2.5$ and $\mu_V=0.05$ ($e_c\simeq 0.724$). The
  vaccine efficiency is set to $(a)$ $e=0.3$, $(b)$ $e=e_c$, and $(c)$ $e=0.9$.}
\label{f4}
\end{figure*}

\subsubsection{Stability of fixed point 4}

As discussed in Sec. \ref{subsec41}, one condition
for stability of this fixed point is $\beta_I<\frac{\beta_V}{\mu_V}$.
In order to discuss the other conditions
we separate again the cases $F>0$ and $F<0$.
Let us define the threshold value $e_c$ for the vaccine efficiency
(see Appendix \ref{appendixA}):
\begin{equation}
{e_c}\equiv \frac{\beta_V }{{1 + \beta_V- \mu_V}}.
\end{equation}

For $F>0$, if $e>e_c$ the fixed point
4 cannot be stable, while if $e<e_c$,
this fixed point is stable and physical in the interval
$\beta_V^* < \beta_I < \beta_E^*$ (see Appendix \ref{appendixB}).
However, for $F<0$, it can be shown that neither for $e>e_c$ nor for $e<e_c$
the fixed point is both physical and stable (see Appendix \ref{appendixC}).
Furthermore, in Appendix~\ref{appendixD} we prove that for $e<e_c$
the parameter $\beta_V^*$ is smaller than $\beta _E^*$ while for $e>e_c$
the opposite is true.

We can summarize the stability of the fixed points
for the case $\beta_V/\mu_V>1$  as follows:
\begin{enumerate}[(1)]
	\item If $e<e_c$, then $ \beta_V^* < \beta _E^* $, and
	\begin{enumerate}[(i)]
		\item For $\beta_I<\beta _V^*$, only fixed point 2 is stable.
		
		\item For $\beta _V^*<\beta_I<\beta _E^*$, only fixed point 4 is stable.
		
		\item For $\beta_I>\beta _E^*$, only fixed point 3 is stable.
	\end{enumerate}
	
	\item If $e>e_c$, then $ \beta_E^* < \beta_V^* $,  and
	\begin{enumerate}[(i)]
		\item For $\beta_I<\beta_E^*$, only fixed point 2 is stable.

		\item For $\beta_E^*<\beta_I<\beta_V^*$, both fixed points 2 and 3 are stable.
		
		\item For $\beta_I>\beta _V^*$, only fixed point 3 is stable.
	\end{enumerate}
\end{enumerate}

These results are confirmed in Figs.~\ref{f2} and \ref{f3} for two
values of vaccine efficiency below and above the threshold value $e_c$.
The stable fixed points are represented by black solid dots in the phase
space $(S,I)$.
Figure~\ref{f2} is plotted for $e=0.3<e_c \simeq 0.724$.
We change the value of
$\beta_I$ such that in Fig.~\ref{f2}(a) the fixed point 2 is
stable, and in Figs.~\ref{f2}(b) and ~\ref{f2}(c) the fixed points 4 and
3 are stable, respectively. Similarly, for $e=0.9>e_c$,
Fig.~\ref{f3} shows the interval of values of $\beta_I$ for which
one or both fixed point 2 or 3 are stable.

\begin{figure}[t]
\begin{center}
\scalebox{0.32}{\includegraphics[angle=0]{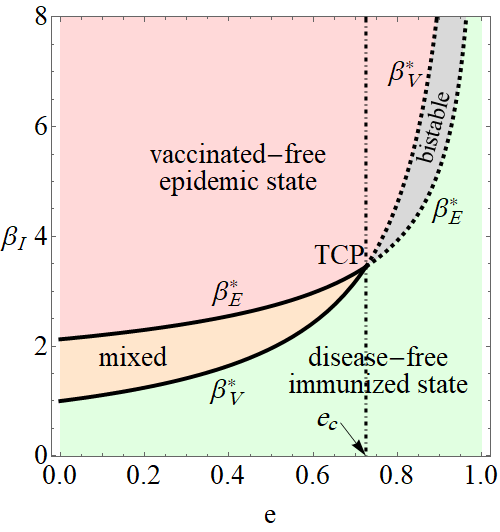}}~~~~
\end{center}
\caption{Phase diagram of the model for $\beta_V/\mu_V>1$
  ($\beta_V=2.5$ , $\mu_V=0.05$) as a function of
  $\beta_I$ and $e$. The solid lines show continuous while dashed
  curves indicate the discontinuous transition points. The tricritical
  point is located at $(e_c,
  \beta_{Ic})=(0.724,3.45)$. The transition lines are
    computed from Eqs.~(\ref{1con}) and~(\ref{2con}).
}
\label{f5}
\end{figure}

The fractions of susceptible (S), infected (I), and vaccinated (V)
nodes are plotted as a function of $\beta_I$ in Fig.~\ref{f4}.
Figure~\ref{f4}(a) shows
the case $e<e_c$. In this case when $\beta_I<\beta _V^*$, we have a
disease-free immunized state and only the fixed point 2 is stable, i.e.,
some nodes are susceptible and the others are vaccinated.
Increasing the value of $\beta_I$, a continuous transition occurs
(fixed point 4 becomes stable),
and for $\beta _V^*<\beta_I<\beta _E^*$ a finite fraction of infected
nodes is present in the stationary state.
As we increase $\beta_I$ further, above
$\beta_I=\beta _E^*$, the fraction of vaccinated nodes becomes zero
(fixed point 3 becomes stable). The transition to this
vaccinated-free epidemic state is also continuous. When $e=e_c$
[Fig.~\ref{f4}(b)], the two transition points $\beta_V^*$ and
$\beta _E^*$ coincide, and a discontinuous transition occurs from the
disease-free immunized state to the vaccinated-free epidemic state.
In other words, at $e=e_c$ the type of the transitions is changed
from continuous to discontinuous and the model exhibits a
tricritical point. For the values we consider, $\beta_V=2.5$ and $\mu_V=0.05$, the
tricritical point occurs at $(e_c,\beta_{Ic})=(0.724,3.45)$.
For $e>e_c$ [Fig.~\ref{f4}(c)], the value of $\beta_E^*$
is smaller than $\beta_V^* $. So, in the interval
$\beta_E^* <\beta_I< \beta_V^*$, both the disease-free immunized and
vaccinated-free epidemic states are possible (both fixed points 2 and 3 are
stable) and bistability emerges. Notice that for $\mu_V >1$,
$e_c$ is larger than 1 and we always have $e<e_c$. Hence the
bistability emerges only for values $\mu_V<1$. We can see the full
phase diagram of the model for the case $\beta_V>\mu_V$ in Fig.~\ref{f5}.

\begin{figure*}[t]
\begin{center}
\scalebox{0.44}{\includegraphics[angle=0]{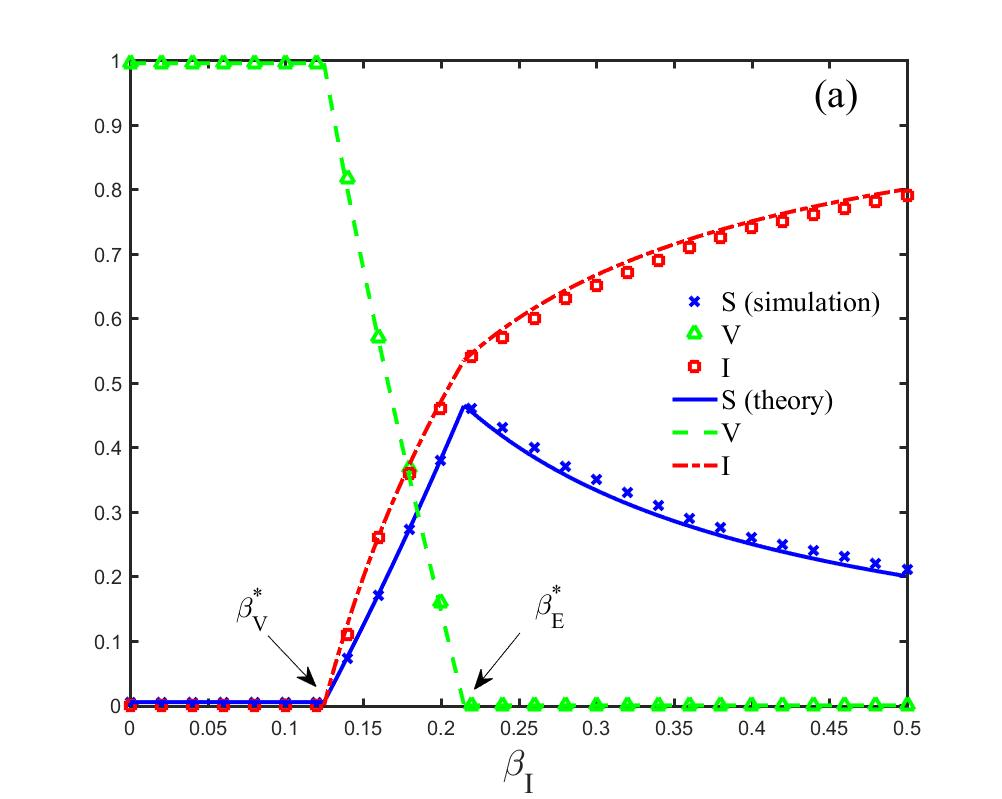}}~~~~
\scalebox{0.44}{\includegraphics[angle=0]{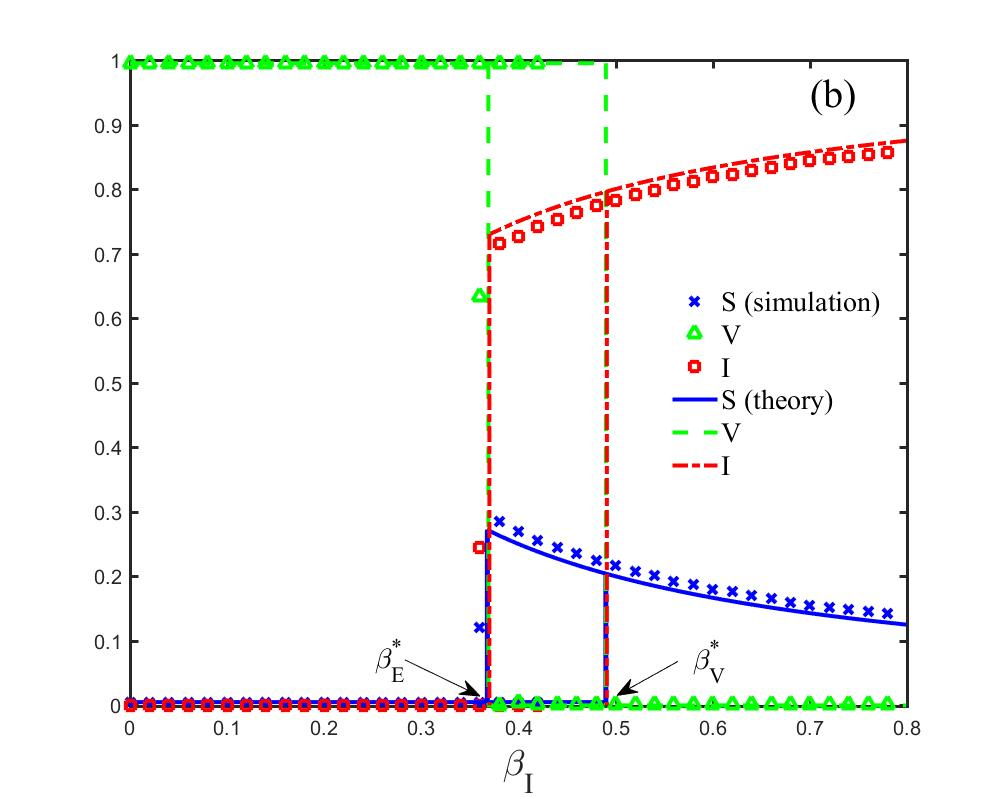}}
\end{center}
\caption{Stationary values for
  susceptible (S), infected (I), and vaccinated (V) densities
  versus the infection rate $\beta_I$ on the $ER$ network with
  $\langle k \rangle=10$. Symbols show the results of numerical
  simulations, which are in agreement with the mean-field approach
  (solid lines). In both panels $\beta_V=0.2$ and
  $\mu_V=0.01$ ($e_c\simeq 0.668$). In $(a)$ the vaccine efficiency is
  set to $e=0.2<e_c$, and the initial condition of the simulations is
  $I_0=0.1$, $V_0=0.9$. Transition points are $\beta_V^*\simeq 0.124$ and
$\beta_E^*\simeq 0.215$.
  In panel $(b)$ $e=0.8>e_c$ and for the forward diagram $I_0=0.1$,$V_0=0.01$,
  while for the backward diagram $I_0=0.01$ and $V_0=0.99$. Transition points are $\beta_E^* \simeq 0.367$ and $\beta_V^*\simeq 0.490$.}
\label{f6}
\end{figure*}
\subsection{The case of perfect vaccination $e=1$}

An interesting special case of the general framework presented above
occurs when the vaccination is fully effective ($e=1$, i.e., $\beta'=0$)
so that the direct transition $V \to I$ is prohibited.
In this case there is a perfect symmetry between the two competing SIS
processes, which is not apparent only because we have set $\mu_I=1$
from the outset, while we have kept $\mu_V$ free. In this case only the
first three fixed points are present: Fixed point 4
disappears for $\beta'=0$. The fixed point 1 is stable if
both $\beta_I<1$ and  $\beta_V/\mu_V<1$.
In such a case both SIS processes spontaneously vanish and the
absorbing, fully susceptible state, is reached.
The other two possible stationary states are fixed point 2
(disease-free immunized state), which is reached if $\beta_V/\mu_V>1$ and
$\beta_V/\mu_V>\beta_I$, and fixed point 3 (vaccinated-free epidemic state) reached if $\beta_I>1$ and $\beta_I > \beta_V/\mu_V$.
We conclude that, in the perfectly symmetric case, only the most
infective SIS process can asymptotically survive, leading to the
complete eradication of the other.
A mixed state with coexistence of I and V individuals
in the stationary state is possible only in the presence
of an imperfect vaccination, i.e., an asymmetry between the two
competing processes.

\section{Numerical Simulations on homogeneous networks}

So far we have considered the mean-field solution of the SIV model,
corresponding to its behavior in the case a node interacts with a
single random neighbor.
A more realistic case is to consider the model on structured networks.
Let us consider the Erd\H{o}s--R\'enyi (ER)
random network with mean degree $\langle k \rangle$, a paradigmatic
example of a homogeneous network.
To describe this system the mean-field equations must be modified,
to take into account that each node is in contact, on average,
with $\langle k \rangle$ other nodes. Hence the equations are
exactly the same provided all transmission rates
($\beta_I,\beta_V,\beta'$) are multiplied by the factor $\langle k \rangle$.
Substituting these values into Eqs.~(\ref{eq1}) and (\ref{eq2}),
the fixed points of the model are now
\begin{eqnarray}
(1) \rightarrow&~~(x_1^*,y_1^*,z_1^*) = &(1,0,0),\\
(2) \rightarrow&~~(x_2^*,y_2^*,z_2^*) = &\left(\frac{\mu_V}{\langle k \rangle\beta_V },0,1 - \frac{\mu_V}{\langle k \rangle\beta_V }\right),~~~ \\
(3) \rightarrow&~~(x_3^*,y_3^*,z_3^*) = &\left(\frac{1}{\langle k \rangle\beta_I},1 -\frac{1}{\langle k \rangle \beta_I},0\right),~~~\\
(4) \rightarrow&~~(x_4^*,y_4^*,z_4^*) = &\bigg(\frac{\mu_V + \langle k \rangle\beta' - 1}{\langle k \rangle(\beta' - \beta_I  + \beta_V )},~~~\nonumber\\
&~~~~~&\frac{\mu_V(\beta_I  - \beta') + \beta_V \left( {\langle k \rangle\beta' - 1} \right)}{\langle k \rangle\beta'\left( {\beta' - \beta_I  + \beta_V } \right)},~~~\nonumber\\
&~~~~~&\frac{\beta'(1-\langle k \rangle\beta_I)+\beta_V-\mu_V\beta_I}{\langle k \rangle\beta'(\beta'-\beta_I+\beta_V)}\bigg).
\end{eqnarray}

All the arguments in the previous section extend to this case,
provided the threshold $e_c$ and the transition points
$\beta_V^*$ and $\beta_E^*$ are redefined as follows:
\begin{eqnarray}
\label{ec}
e_c &=& \frac{\langle k \rangle\beta_V }{{1 + \langle k \rangle\beta_V  - \mu_V}},\\
\beta_V^* &=&\frac{\beta_V }{\langle k \rangle\beta_V(1-e)+e\mu_V},\\
\beta_E^*&=& \frac{(1-\mu_V-e) + \sqrt{(\mu_V+e-1)^2 + 4\langle k \rangle\beta_V(1-e)}} {2(1-e)\langle k \rangle}.\nonumber\\
\label{eq16}
\end{eqnarray}
As we can see from Eq.~(\ref{ec}), the critical value of
vaccine efficiency depends on the connectivity $\av{k}$ of the network.
When the connectivity is increased, $e_c$ approaches 1 and so the
region of bistability tends to disappear.
This observation can be rationalized as follows. For a strongly leaky
vaccine, $e < e_c$, a mixed state arises due to the presence of a loop
of transitions ($S \to V \to I \to S$),
in a way analogous to the rock-paper-scissors dynamics \cite{RSP}.
Bistability is observed only when the $V \to I$ process in this
loop is suppressed, which happens when $\beta'$ is sufficiently small
(i.e., the vaccine efficiency is larger than the critical value $e_c$).
Increasing $\av{k}$ implies that vaccinated individuals have more
infected neighbors and so their chance of getting infected is increased:
in other words, the process $V \to I$ is enhanced. In order to see
bistability one needs $\beta'$ to be reduced to compensate for the increase
of connectivity. This explains why $e_c$ grows with $\av{k}$ and
bistability tends to disappear as connectivity becomes large.

To validate the analytical results,
we perform (using a continuous time Gillespie algorithm),
numerical simulations of the SIV dynamics
on ER random networks.
We consider a network consisting of
$N=10^4$ nodes and with mean degree $\langle k \rangle=10$ and select
parameters of the model corresponding to the two cases $e<e_c$ and
$e>e_c$.
Let us set $\beta_V=0.2$ and $\mu_V=0.01$, so that $\beta_V/\mu_V>1$.
According to Eq.~(\ref{eq16}), the
efficiency threshold is $e_c\simeq 0.668$. We first consider the
value $e=0.2 < e_c$ so that we expect the presence of a mixed
state.
We choose the initial conditions as
$I_0=0.1$, $V_0=0.9$ and $S_0=0$ and average over 20 realizations.
Figure.~\ref{f6}(a) shows the
stationary values for the fractions of susceptible, infected, and
vaccinated nodes as a function of the infection rate $\beta_I$.
We can see that numerical simulations (symbols) are in good agreement with
analytical results (curves), obtained in the previous sections. In this
case the transition points are $\beta_V^*\simeq 0.124$ and
$\beta_E^*\simeq 0.215$ [Eq.~(\ref{eq16})], which are very close to the
numerical results.
Next we consider a vaccine efficiency $e=0.8>e_c$.
In this case the dependence of the densities on $\beta_I$
is qualitatively different and a hysteresis loop appears
[Fig.~\ref{f6}(b)].
For the initial condition $I_0=0.1$ and $V_0=0.01$ a discontinuous
transition occurs for $\beta_I \simeq \beta_V^*\simeq 0.490$.
If we choose the initial condition as $I_0=0.01$ and $V_0=0.99$,
the transition point is instead $\beta_I \simeq \beta_E^* \simeq 0.367$.

\section{Conclusions}

In this work, we have studied a three-state SIV model in which disease
spreading competes with a vaccination process. We have assumed both
disease and vaccination spreading follow the dynamics of the standard
SIS model. Hence, similar to the disease, the vaccination is also
considered as a contact process such that vaccinated individuals
convince their susceptible neighbors to be vaccinated. We have
assumed an imperfect vaccination so that vaccinated individuals
can be infected at a reduced rate. This couples
asymmetrically the two competing models.

We have identified the existence of two completely different scenarios.
If the vaccination rate is not large enough with respect to the rate
at which immunity fades away, the vaccination process does not have any
effect on disease spreading in the stationary state and the same
phase diagram of the standard SIS model is obtained.
Instead, if the vaccination is large enough new
``disease-free immunized" and ``mixed" phases appear.
For a small value of vaccine efficiency, the model shows two continuous
transitions as the infection rate is increased. The first transition
occurs from the disease-free immunized phase,
in which only susceptible and vaccinated nodes exist,
to the mixed phase with a mixture of susceptible,
infected, and vaccinated nodes.
The second transition occurs at a higher infection rate and separates
the mixed phase from the vaccinated-free epidemic phase, in which
no vaccinated nodes are present.
For larger vaccine efficiency, above a tricritical point,
the mixed phase disappears and is replaced by a bistability region,
with both disease-free immunized and vaccinated-free epidemic states stable.

We have checked that the MF scenario depicted above is observed also
when the interaction pattern is described by a homogeneous network.
Whether this remains true also for more complex topologies (such as
heterogeneous, clustered, or correlated networks) is a promising
avenue for future research.
Another path that could be followed is the investigation
of the role of the imperfect vaccination.
Its presence creates an asymmetric direct coupling between V and I states,
that induces the possibility of cyclic transitions in the model.
It would be interesting to analyze the similarities and differences
with respect to other cyclical competing three-state dynamics,
such as the rock-paper-scissors model.

\appendix
\section{}
\label{appendixA}

For $F=\beta_V+\beta'-\beta_I <0$, from the physical condition $ 0<x_4^*<1 $ we obtain that $ \beta_I$ must be less than $\frac{1-\mu_V}{1-e} $ and greater than $1+\beta_V-\mu_V $. An overlap interval exists if the following subtraction is positive:
\begin{equation}
\frac{1-\mu_V}{1-e} -(1+\beta_V-\mu_V)= \frac{e-\beta_V+e\beta_V-e\mu_V}{1-e}.
\label{A1}
\end{equation}
Let us define a threshold value for efficiency of vaccination as follows:
\begin{equation}
{e_c}\equiv \frac{\beta_V }{{1 + \beta_V- \mu_V}}.
\label{A2}
\end{equation}
The relation (\ref{A1}) is positive if $e>e_c$. In this case the physical condition for $x_4^*$ is satisfied. On the other hand, from  $ y_4^*>0 $ it is concluded that $ \beta_I$ also must be less than  $\frac{\beta_V }{\beta_V(1-e)+e\mu_V}\equiv\beta_V^*$. That means $\beta_V^*$ must be greater than $1+\beta_V-\mu_V$ or the following subtraction must be negative:
\begin{equation}
\begin{aligned}
&(1+\beta_V-\mu_V) -\beta_V^*  =(1+\beta_V-\mu_V)  \\
& -\frac{\beta_V}{\beta_V(1-e)+e\mu_V}\\
& = \frac{(e-\beta_V+e\beta_V-e\mu_V)(\mu_V-\beta_V)}{\beta_V(1-e)+e\mu_V}.
\end{aligned}
\label{A3}
\end{equation}
For $e>e_c$, we have $(e+e\beta_V-e\mu_V - \beta_V) > 0$, and since $\mu_V> \beta_V$, the subtraction is always positive. Hence the physical conditions for $x_4^*$ and $y_4^*$ are not satisfied simultaneously.

\section{}
\label{appendixB}

For $ F>0 $, as we discussed in Subsec.~(\ref{subsec41}), the stability condition for the fourth fixed point leads to Eq.~(\ref{eq7}). If one substitutes $\beta'=\beta_I(1-e)$, Eq.~(\ref{eq7}) can be rewritten as an inequation of order 3 for $\beta_I$:
\begin{equation}
g(\beta_I)\equiv a\beta_I^3+b\beta_I^2+c\beta_I+d>0,
\label{beta1}
\end{equation}
where
\begin{equation}
a= -(1-e)\left(e\mu_V+\beta_V(1-e)\right) <0
\label{b3equ}
\end{equation}
and $b$, $c$, and $d$ are some parameters. $g(\beta_I)=0$ has three roots:
\begin{equation}
{\beta _{I1}} = \frac{{(1 - \mu_V - e) + \sqrt {{{(\mu_V + e - 1)}^2} + 4\beta_V (1 - e)} }}{{2(1 - e)}}\equiv\beta _E^*,
\end{equation}
\begin{equation}
{\beta _{I2}} = \frac{{(1 - \mu_V - e) - \sqrt {{{(\mu_V + e - 1)}^2} + 4\beta_V (1 - e)} }}{{2(1 - e)}},
\end{equation}
\begin{equation}
{\beta _{I3}} =\frac{\beta_V }{{\beta_V(1-e) + e\mu_V }}\equiv\beta _V^*,
\label{roots}
\end{equation}
such that $\beta _E^*$ and $\beta _V^*$ are positive but $\beta_{I2}$
is negative. Since the coefficient of cubic term (parameter $a$) is
always negative, we conclude that the inequality (\ref{beta1}) is
satisfied for $ \beta_I $ between positive roots $\beta_E^*$ and
$\beta_V^*$. On the other hand, $\beta_I$ must satisfy the two additional
conditions $\beta_I<\frac{\beta_V}{\mu_V}$ and
$\beta_I<\frac{\beta_V}{e}$. Let us consider two cases:
\begin{enumerate}[(1)]
	\item $e>e_c$: According to Appendix \ref{appendixF}, both $\beta_E^*$ and $\beta_V^*$ are greater than $\frac{\beta_V}{e}$, Hence the condition $\beta_I<\frac{\beta_V}{e}$ is not satisfied.
	\item $e<e_c$: In this case, $\beta_V^*$ is less than $\beta_E^*$ (see Appendix \ref{appendixD}), and both are less than $\frac{\beta_V}{\mu_V}$ and $\frac{\beta_V}{e}$ as proved in Appendixes \ref{appendixE} and \ref{appendixF}, respectively.
\end{enumerate}

Hence, the fourth fixed point is stable in the interval $\beta_V^*<\beta_I<\beta_E^*$ when $e<e_c$ and $F>0$. Now, we must check the following six physical conditions:
 	
\begin{enumerate}[(1)]
 		\item $ x_4^*>0 \to \beta_I >\frac{1-\mu_V}{1-e} $
 		\item $ x_4^*<1 \to \beta_I <1+\beta_V-\mu_V$ 
 		\item $ y_4^*>0 \to \beta_I > \beta_V^* $
 		\item $ y_4^*<1 \to \beta_I < \beta_{c1}\\ 
 		\beta_{c1}\equiv\frac{-e\mu_V+\sqrt{e^2\mu_V^2+4\beta_V e(1-e)}}{2e(1-e)} $	
 		\item $z_4^*>0  \to \beta_I < \beta_E^* $ 
 		\item $z_4^*<1  \to \beta_I > \beta_{c2} \\
 		\resizebox{.82\hsize}{!}{$ \beta_{c2}\equiv \frac{{(1-\beta_V)(1-e)-\mu_V + \sqrt{{((1-\beta_V)(1-e)-\mu_V)^2} + 4\beta_V(1-e)^2} }}{{2(1-e)^2}}  $ } $
 	\end{enumerate}

Let us specify whether they can be satisfied or not, one by one:
\begin{enumerate}[(1)]
		\item Subtracting $ \frac{1-\mu_V}{1-e} $ from $ \beta_V^* $ (Appendix \ref{appendixG}), we conclude $ \beta_V^*$ is more than it, so the condition is satisfied.
		
		\item The result of subtracting $ \beta_E^* $ from $ (1+\beta_V-\mu_V) $ is positive (Appendix \ref{appendixH}), hence this condition is true as well.
		
		\item This one is clearly correct.
		
		\item After some algebra, it is proved that for this state, $ \beta_E^* $ is less than $\beta_{c1} $ (Appendix \ref{appendixI}), so this condition is also satisfied.
		
		\item Obviously, that's correct.
		
		\item According to Appendix \ref{appendixJ}, $\beta_{c2} < \beta_V^* $, so there is no problem with this condition too.
\end{enumerate}

So far we conclude that the fourth fixed point is stable and physical for the interval  $ \beta_V^* < \beta_I < \beta_E^* $ when $ e<e_c $.

\section{}
\label{appendixC}
If $ F<0$ we must have $ X < -F(\mu_V\beta_I-\beta_V)^2 $, such that the sign of Eq.~(\ref{eq7}) is reversed. Substituting $\beta'=\beta_I(1-e)$, we obtain
\begin{equation}
g(\beta_I)\equiv a\beta_I^3+b\beta_I^2+c\beta_I+d<0
\label{beta3}
\end{equation}
where $ a $ is the same as (\ref{b3equ}). In addition, from $ F<0 $ and the condition $\beta_I <\frac{\beta_V}{\mu_V }$, we find that $ \frac{\beta_V}{e} < \beta_I < \frac{\beta_V}{\mu_V}$, which results in $ \mu_V<e<1 $. Let us check both stability and physical conditions for the obtained interval in two cases:
 	
\begin{enumerate}[(1)]
\item $ e<e_c $: According to Appendix \ref{appendixF}, both $ \beta_V^* $ and $ \beta_E^* $, are less than $ \frac{\beta_V}{e} $. Then $g(\beta_I)$ is always negative. Hence, the condition $g(\beta_I)<0$ is satisfied. The condition $ x_4^*>0$, leads to $ \beta_I <\frac{1-\mu_V}{1-e} $. On the other hand, $ \frac{1-\mu_V}{1-e} < \frac{\beta_V}{e} $. Therefore, stability and the physical condition are not satisfied at the same time.

\item $ e>e_c $: In this case both $ \beta_V^* $ and $ \beta_E^* $ are greater than $\frac{\beta_V}{e} $. From Appendixes D and E, we have $ \beta_E^* < \beta_V^*< \frac{\beta_V}{\mu_V}$. So, the fourth fixed point can be stable in the intervals $ \frac{\beta_V}{e} <\beta_I <  \beta_E^* $ and  $ \beta_V^* < \beta_I < \frac{\beta_V}{\mu_V} $. On the other hand, for $ y_4^*>0 $ we need $ \beta_I < \beta_V^*$, and for $ z_4^*>0 $ we must have $ \beta_I>\beta_E^* $, which obviously contradict obtained intervals. So the fourth fixed point is not physical in this case as well.
\end{enumerate}

\section{}
\label{appendixD}


Let us assume that $\beta_E^* - \beta_V^*> 0$. In this case we have the following inequality:
\begin{eqnarray}
&&\frac{{(1 - \mu_V - e) + \sqrt {{{(\mu_V + e - 1)}^2} + 4\beta_V (1 - e)} }}{{2(1 - e)}}\nonumber\\
 &&- \frac{\beta_V}{\beta_V(1-e)+e\mu_V} > 0.
\label{D2}
\end{eqnarray}
The denominator of inequality (\ref{D2}) is positive. Hence the numerator must be positive as well:
\begin{eqnarray}
&&\sqrt {{{(\mu_V + e - 1)}^2} + 4\beta_V (1 - e)}> \nonumber\\
&& \frac{2\beta_V(1-e)}{\beta_V(1-e)+e\mu_V} - (1-e-\mu_V).
\end{eqnarray}
After some calculations, we obtain that
\begin{eqnarray}
&&e^2(\beta_V-\mu_V)(\beta_V -\mu_V +1)\nonumber\\
&&-e(\beta_V-\mu_V)(\beta_V-\mu_V+\beta_V+1)\nonumber\\
&&+\beta_V(\beta_V-\mu_V) > 0.
\end{eqnarray}
Notice that for case $ \beta_V>\mu_V $, the coefficient of $ e^2 $ and the third term are both positive. The discriminant of the related quadratic equation is $ \Delta = (\beta_V-\mu_V)^2 (\mu_V-1)^2 $, which is also positive. Therefore, there are two real roots, $ e_1=\dfrac{\beta_V}{1+\beta_V-\mu_V}=e_c $ and $ e_2=1 $. These roots are both positive for the case $ \beta_V>\mu_V $. Regarding to the value of $\mu_V $, there are two statuses:
\begin{enumerate}[(1)]
	\item If $ \mu_V<1 $, then $ e_1<e_2 $. In this case for $e<e_c$ we have $\beta_E^* > \beta_V^*$, and for the opposite result is obtained for $ e>e_c $.
	\item If $ \mu_V>1 $, then $ e_1>e_2 $ and for each value of $e$, we have $\beta_E^* > \beta_V^*$. However, $e_1>e_2$ means that $e_c>1$, which never occurs since efficiency is always less than 1.
\end{enumerate}

\section{}
\label{appendixE}
In order to compare $\beta_V^*$ and $\frac{\beta_V}{\mu_V}$, we discuss the sign of their difference:
\begin{eqnarray}
\beta_V^* - \frac{\beta_V}{\mu_V}=&\frac{\beta_V}{\beta_V(1-e)+e\mu_V} - \frac{\beta_V}{\mu_V}\nonumber\\
=&\frac{\beta_V(\mu_V-\beta_V)(1-e)}{\mu_V(\beta_V(1-e)+e\mu_V))}.
\end{eqnarray}
If $\beta_V>\mu_V$, it is concluded that $\beta_V^* < \frac{\beta_V}{\mu_V}$ and in the opposite case $ \beta_V<\mu_V $, we obtain the opposite result.
%

Similarly, we consider the subtraction $\beta_E^* - \frac{\beta_V}{\mu_V}$:
\begin{eqnarray}
&&\beta_E^* - \frac{\beta_V}{\mu_V}=\nonumber\\
&&\frac{{(1 - \mu_V - e) + \sqrt {{{(\mu_V + e - 1)}^2} + 4\beta_V (1 - e)} }}{{2(1 - e)}}\nonumber\\
&& - \frac{\beta_V}{\mu_V}.
\end{eqnarray}

Suppose that the result of this subtraction is negative. Since, the denominator is positive, it is concluded that 
\begin{eqnarray}
&&\mu_V\sqrt {{{(\mu_V + e - 1)}^2} + 4\beta_V (1 - e)} < \nonumber\\
&&2\beta_V(1-e)-\mu_V(1-e-\mu_V).
\end{eqnarray}
After some calculations, it is obtained that $ (\beta_V-\mu_V)(e-1) < 0 $,  which is always correct for the case $ \mu_V<\beta_V $. Therefore, we conclude that
\begin{enumerate}[(i)]
	\item If  $ \beta_V>\mu_V $, then $  \beta_E^* < \frac{\beta_V}{\mu_V} $.
	
	\item If  $ \beta_V<\mu_V $, then $  \beta_E^* > \frac{\beta_V}{\mu_V} $.
\end{enumerate}

\section{}
\label{appendixF}
Let us consider the following subtraction:
\begin{eqnarray}
\beta_V^* - \frac{\beta_V}{e}=&\frac{\beta_V}{\beta_V(1-e)+e\mu_V} - \frac{\beta_V}{e}\nonumber\\
=&\frac{\beta_V(e-e\mu_V-\beta_V+e\beta_V)}{e(\beta_V(1-e)+e\mu_V)}.
\end{eqnarray}
Since the denominator is positive, for $e>e_c$ we conclude that $(e+e\beta_V-e\mu_V - \beta_V) > 0$ and therefore $\beta_V^* > \frac{\beta_V}{e}$. In the same way for $e<e_c$, we obtain the opposite result, $\beta_V^* < \frac{\beta_V}{e}$.

Also we calculate the following subtraction:
\begin{eqnarray}
&&\beta_E^* - \frac{\beta_V}{e}=\nonumber\\
&&\frac{{(1 - \mu_V - e) + \sqrt {{{(\mu_V + e - 1)}^2} + 4\beta_V (1 - e)} }}{{2(1 - e)}} - \frac{\beta_V}{e}\nonumber\\
&&=\frac{{e(1 - \mu_V - e) +e \sqrt {{{(\mu_V + e - 1)}^2} + 4\beta_V (1 - e)}  }}{{2e(1 - e)}}\nonumber\\
&&-\frac{2\beta_V(1-e)}{2e(1-e)}.
\end{eqnarray}
Let us assume that the result is positive. The denominator is positive, and therefore the nominator must be positive:
\begin{eqnarray}
&&e\sqrt {{{(\mu_V + e - 1)}^2} + 4\beta_V (1 - e)} >\nonumber\\
&&2\beta_V(1-e)-e(1-e-\mu_V).
\end{eqnarray}
After some calculations, we obtain that $ (e+e\beta_V-e\mu_V - \beta_V) > 0$,  which leads to  $e>e_c $. Consequently, for $e>e_c$ we get that $\beta_E^* > \frac{\beta_V}{e}$ and for $e<e_c$ the opposite result, namely, $\beta_E^* < \frac{\beta_V}{e}$, is obtained.

\section{}
\label{appendixG}

We can easily obtain that
\begin{eqnarray}
\beta_V^* - \frac{1-\mu_V}{1-e}=&&\frac{\beta_V}{\beta_V(1-e)+e\mu_V} - \frac{1-\mu_V}{1-e}\nonumber\\
=&&\frac{\mu_V(\beta_V-e-e\beta_V+e\mu_V)}{(\beta_V(1-e)+e\mu_V)(1-e)}.
\end{eqnarray}
Since the denominator is positive, in the case that $ e>e_c$ we obtain $(e+e\beta_V-e\mu_V - \beta_V)>0$. It results in
\begin{eqnarray}
	\beta_V^* < \frac{1-\mu_V}{1-e}.
\end{eqnarray}
With the same argument for $e<e_c$, the opposite result is obtained.

Also, we can see the difference between $\beta_E^*$ and $\frac{1-\mu_V}{1-e}$:
\begin{eqnarray}
&&\beta_E^* - \frac{1-\mu_V}{1-e}=\nonumber\\
&&\frac{{(1 - \mu_V - e) + \sqrt {{{(\mu_V + e - 1)}^2} + 4\beta_V (1 - e)} }}{{2(1 - e)}}\nonumber \\
&&- \frac{1-\mu_V}{1-e}.
\end{eqnarray}
If we assume that the result is negative, then since the denominator is positive we get
\begin{eqnarray}
\sqrt {{{(\mu_V + e - 1)}^2} + 4\beta_V (1 - e)} < 1+e-\mu_V.
\end{eqnarray}
Hence, we conclude that $ e+e\beta_V-e\alpha -\beta_V > 0 $, which is true for the case $ e>e_c $. Therefore for  $e>e_c$ it is obtained that $\beta_E^* < \frac{1-\mu_V}{1-e}$, and for the case $e<e_c $ the sign of inequality is opposite and $\beta_E^*$ is greater than $\frac{1-\mu_V}{1-e}$.

\section{}
\label{appendixH}

Next we consider the following subtraction:
\begin{eqnarray}
&&\beta_E^* -(1+\beta_V-\mu_V)=\nonumber\\
&&\frac{{(1 - \mu_V - e) + \sqrt {{{(\mu_V + e - 1)}^2} + 4\beta_V (1 - e)} }}{{2(1 - e)}}\nonumber \\
&&-(1+\beta_V-\mu_V).
\end{eqnarray}
If the result is negative, since the denominator is positive we can write
\begin{eqnarray}
&&\sqrt {{{(\mu_V + e - 1)}^2} + 4\beta_V (1 - e)} < \nonumber\\
&&e+\mu_V-1+2(1-e)(1+\beta_V-\mu_V).
\end{eqnarray}
Then $ (\beta_V-\mu_V)(\beta_V-e\beta_V+e\mu_V-e) >0 $ if $ \beta_V >\mu_V$ and $e<e_c $. Hence we can summarize the results as follows:
\begin{enumerate}[(1)]
	\item  For  $ \beta_V >\mu_V$ :
	\begin{enumerate}[(i)]
		\item If $ e<e_c $ then $\beta_E^* < (1+\beta_V-\mu_V) $
		\item If $ e>e_c $ then $\beta_E^* > (1+\beta_V-\mu_V) $
	\end{enumerate}
	\item For $ \beta_V <\mu_V$ :
	\begin{enumerate}
		\item If $ e<e_c $ then $\beta_E^* > (1+\beta_V-\mu_V) $
		\item If $ e>e_c $ then $\beta_E^* < (1+\beta_V-\mu_V) $
	\end{enumerate}

\end{enumerate}


\section{}
\label{appendixI}
We assume that the following subtraction has positive sign:
\begin{eqnarray}
&&\beta_{c1} - \beta_E^*=\frac{-e\mu_V+\sqrt{e^2\mu_V^2+4\beta_V e(1-e)}}{2e(1-e)}\nonumber\\
&&-\frac{{(1 - \mu_V - e) + \sqrt {{{(\mu_V + e - 1)}^2} + 4\beta_V (1 - e)} }}{2(1-e)} >0.\nonumber\\
\end{eqnarray}
The denominator is positive, hence we obtain $ \frac{4\beta_V(1-e)(e\mu_V-e-e\beta_V+\beta_V)}{e^2} >0 $. This result is satisfied when $ e<e_c $. Therefore, for $ e>e_c $ we conclude that $\beta_{c1} < \beta_E^*$, while for $e<e_c$, the opposite result, namely,  $\beta_{c1}>\beta_E^*$ is correct.

\section{}
\label{appendixJ}
Let us assume that $\beta_V^*$ is greater than $\beta_{c2}$, such that the sign of following subtraction is positive:
\begin{eqnarray}
&&\beta_V^* - \beta_{c2} = \frac{\beta_V}{\beta_V(1-e)+e\mu_V}\nonumber\\
&&- \frac{{(1-\beta_V)(1-e)-\mu_V }}{{2(1-e)^2}}\nonumber\\
&&- \frac{\sqrt{{((1-\beta_V)(1-e)-\mu_V)^2} + 4\beta_V(1-e)^2}}{2(1-e)^2}.
\end{eqnarray}
The denominator is positive, hence we can write: 
\begin{eqnarray}
-e(1-e)[(\mu_V-\beta_V)^2+(\beta_V^2-\mu_V)]
-\mu_V\beta_V(1+e^2)<0.\nonumber\\
\end{eqnarray}
which is always correct for the case $ \beta_V>\mu_V $. In other words if $ \beta_V>\mu_V$, we have $ \beta_V^*> \beta_{c2}$.

\bibliography{vaccine-SIS}

\begin{thebibliography}{33}%
\makeatletter
\providecommand \@ifxundefined [1]{%
 \@ifx{#1\undefined}
}%
\providecommand \@ifnum [1]{%
 \ifnum #1\expandafter \@firstoftwo
 \else \expandafter \@secondoftwo
 \fi
}%
\providecommand \@ifx [1]{%
 \ifx #1\expandafter \@firstoftwo
 \else \expandafter \@secondoftwo
 \fi
}%
\providecommand \natexlab [1]{#1}%
\providecommand \enquote  [1]{``#1''}%
\providecommand \bibnamefont  [1]{#1}%
\providecommand \bibfnamefont [1]{#1}%
\providecommand \citenamefont [1]{#1}%
\providecommand \href@noop [0]{\@secondoftwo}%
\providecommand \href [0]{\begingroup \@sanitize@url \@href}%
\providecommand \@href[1]{\@@startlink{#1}\@@href}%
\providecommand \@@href[1]{\endgroup#1\@@endlink}%
\providecommand \@sanitize@url [0]{\catcode `\\12\catcode `\$12\catcode
  `\&12\catcode `\#12\catcode `\^12\catcode `\_12\catcode `\%12\relax}%
\providecommand \@@startlink[1]{}%
\providecommand \@@endlink[0]{}%
\providecommand \url  [0]{\begingroup\@sanitize@url \@url }%
\providecommand \@url [1]{\endgroup\@href {#1}{\urlprefix }}%
\providecommand \urlprefix  [0]{URL }%
\providecommand \Eprint [0]{\href }%
\providecommand \doibase [0]{http://dx.doi.org/}%
\providecommand \selectlanguage [0]{\@gobble}%
\providecommand \bibinfo  [0]{\@secondoftwo}%
\providecommand \bibfield  [0]{\@secondoftwo}%
\providecommand \translation [1]{[#1]}%
\providecommand \BibitemOpen [0]{}%
\providecommand \bibitemStop [0]{}%
\providecommand \bibitemNoStop [0]{.\EOS\space}%
\providecommand \EOS [0]{\spacefactor3000\relax}%
\providecommand \BibitemShut  [1]{\csname bibitem#1\endcsname}%
\let\auto@bib@innerbib\@empty
\bibitem [{\citenamefont {Newman}(2002)}]{Newman}%
  \BibitemOpen
  \bibfield  {author} {\bibinfo {author} {\bibfnamefont {M.~E.~J.}\
  \bibnamefont {Newman}},\ }\href {\doibase 10.1103/PhysRevE.66.016128}
  {\bibfield  {journal} {\bibinfo  {journal} {Phys. Rev. E}\ }\textbf {\bibinfo
  {volume} {66}},\ \bibinfo {pages} {016128} (\bibinfo {year}
  {2002})}\BibitemShut {NoStop}%
\bibitem [{\citenamefont {Pastor-Satorras}\ \emph {et~al.}(2015)\citenamefont
  {Pastor-Satorras}, \citenamefont {Castellano}, \citenamefont {Van~Mieghem},\
  and\ \citenamefont {Vespignani}}]{castellano}%
  \BibitemOpen
  \bibfield  {author} {\bibinfo {author} {\bibfnamefont {R.}~\bibnamefont
  {Pastor-Satorras}}, \bibinfo {author} {\bibfnamefont {C.}~\bibnamefont
  {Castellano}}, \bibinfo {author} {\bibfnamefont {P.}~\bibnamefont
  {Van~Mieghem}}, \ and\ \bibinfo {author} {\bibfnamefont {A.}~\bibnamefont
  {Vespignani}},\ }\href {\doibase 10.1103/RevModPhys.87.925} {\bibfield
  {journal} {\bibinfo  {journal} {Rev. Mod. Phys.}\ }\textbf {\bibinfo {volume}
  {87}},\ \bibinfo {pages} {925} (\bibinfo {year} {2015})}\BibitemShut
  {NoStop}%
\bibitem [{\citenamefont {Wang}\ \emph {et~al.}(2016)\citenamefont {Wang},
  \citenamefont {Bauch}, \citenamefont {Bhattacharyya}, \citenamefont
  {d'Onofrio}, \citenamefont {Manfredi}, \citenamefont {Perc}, \citenamefont
  {Perra}, \citenamefont {Salathé},\ and\ \citenamefont
  {Zhao}}]{vaccine-review}%
  \BibitemOpen
  \bibfield  {author} {\bibinfo {author} {\bibfnamefont {Z.}~\bibnamefont
  {Wang}}, \bibinfo {author} {\bibfnamefont {C.~T.}\ \bibnamefont {Bauch}},
  \bibinfo {author} {\bibfnamefont {S.}~\bibnamefont {Bhattacharyya}}, \bibinfo
  {author} {\bibfnamefont {A.}~\bibnamefont {d'Onofrio}}, \bibinfo {author}
  {\bibfnamefont {P.}~\bibnamefont {Manfredi}}, \bibinfo {author}
  {\bibfnamefont {M.}~\bibnamefont {Perc}}, \bibinfo {author} {\bibfnamefont
  {N.}~\bibnamefont {Perra}}, \bibinfo {author} {\bibfnamefont
  {M.}~\bibnamefont {Salathé}}, \ and\ \bibinfo {author} {\bibfnamefont
  {D.}~\bibnamefont {Zhao}},\ }\href {\doibase
  https://doi.org/10.1016/j.physrep.2016.10.006} {\bibfield  {journal}
  {\bibinfo  {journal} {Physics Reports}\ }\textbf {\bibinfo {volume} {664}},\
  \bibinfo {pages} {1 } (\bibinfo {year} {2016})},\ \bibinfo {note}
  {statistical physics of vaccination}\BibitemShut {NoStop}%
\bibitem [{\citenamefont {Pastor-Satorras}\ and\ \citenamefont
  {Vespignani}(2002)}]{Pastor}%
  \BibitemOpen
  \bibfield  {author} {\bibinfo {author} {\bibfnamefont {R.}~\bibnamefont
  {Pastor-Satorras}}\ and\ \bibinfo {author} {\bibfnamefont {A.}~\bibnamefont
  {Vespignani}},\ }\href {\doibase 10.1103/PhysRevE.65.036104} {\bibfield
  {journal} {\bibinfo  {journal} {Phys. Rev. E}\ }\textbf {\bibinfo {volume}
  {65}},\ \bibinfo {pages} {036104} (\bibinfo {year} {2002})}\BibitemShut
  {NoStop}%
\bibitem [{\citenamefont {Schneider}\ \emph {et~al.}(2012)\citenamefont
  {Schneider}, \citenamefont {Mihaljev},\ and\ \citenamefont
  {Herrmann}}]{Inverse}%
  \BibitemOpen
  \bibfield  {author} {\bibinfo {author} {\bibfnamefont {C.~M.}\ \bibnamefont
  {Schneider}}, \bibinfo {author} {\bibfnamefont {T.}~\bibnamefont {Mihaljev}},
  \ and\ \bibinfo {author} {\bibfnamefont {H.~J.}\ \bibnamefont {Herrmann}},\
  }\href {\doibase 10.1209/0295-5075/98/46002} {\bibfield  {journal} {\bibinfo
  {journal} {{EPL} (Europhysics Letters)}\ }\textbf {\bibinfo {volume} {98}},\
  \bibinfo {pages} {46002} (\bibinfo {year} {2012})}\BibitemShut {NoStop}%
\bibitem [{\citenamefont {Wang}\ \emph {et~al.}(2017)\citenamefont {Wang},
  \citenamefont {Moreno}, \citenamefont {Boccaletti},\ and\ \citenamefont
  {Perc}}]{Wang}%
  \BibitemOpen
  \bibfield  {author} {\bibinfo {author} {\bibfnamefont {Z.}~\bibnamefont
  {Wang}}, \bibinfo {author} {\bibfnamefont {Y.}~\bibnamefont {Moreno}},
  \bibinfo {author} {\bibfnamefont {S.}~\bibnamefont {Boccaletti}}, \ and\
  \bibinfo {author} {\bibfnamefont {M.}~\bibnamefont {Perc}},\ }\href {\doibase
  https://doi.org/10.1016/j.chaos.2017.06.004} {\bibfield  {journal} {\bibinfo
  {journal} {Chaos, Solitons \& Fractals}\ }\textbf {\bibinfo {volume} {103}},\
  \bibinfo {pages} {177 } (\bibinfo {year} {2017})}\BibitemShut {NoStop}%
\bibitem [{\citenamefont {Schneider}\ \emph {et~al.}(2011)\citenamefont
  {Schneider}, \citenamefont {Mihaljev}, \citenamefont {Havlin},\ and\
  \citenamefont {Herrmann}}]{target}%
  \BibitemOpen
  \bibfield  {author} {\bibinfo {author} {\bibfnamefont {C.~M.}\ \bibnamefont
  {Schneider}}, \bibinfo {author} {\bibfnamefont {T.}~\bibnamefont {Mihaljev}},
  \bibinfo {author} {\bibfnamefont {S.}~\bibnamefont {Havlin}}, \ and\ \bibinfo
  {author} {\bibfnamefont {H.~J.}\ \bibnamefont {Herrmann}},\ }\href {\doibase
  10.1103/PhysRevE.84.061911} {\bibfield  {journal} {\bibinfo  {journal} {Phys.
  Rev. E}\ }\textbf {\bibinfo {volume} {84}},\ \bibinfo {pages} {061911}
  (\bibinfo {year} {2011})}\BibitemShut {NoStop}%
\bibitem [{\citenamefont {Cohen}\ \emph {et~al.}(2003)\citenamefont {Cohen},
  \citenamefont {Havlin},\ and\ \citenamefont {ben Avraham}}]{Acquaintance}%
  \BibitemOpen
  \bibfield  {author} {\bibinfo {author} {\bibfnamefont {R.}~\bibnamefont
  {Cohen}}, \bibinfo {author} {\bibfnamefont {S.}~\bibnamefont {Havlin}}, \
  and\ \bibinfo {author} {\bibfnamefont {D.}~\bibnamefont {ben Avraham}},\
  }\href {\doibase 10.1103/PhysRevLett.91.247901} {\bibfield  {journal}
  {\bibinfo  {journal} {Phys. Rev. Lett.}\ }\textbf {\bibinfo {volume} {91}},\
  \bibinfo {pages} {247901} (\bibinfo {year} {2003})}\BibitemShut {NoStop}%
\bibitem [{\citenamefont {Gandon}\ \emph {et~al.}(2001)\citenamefont {Gandon},
  \citenamefont {Mackinnon}, \citenamefont {Nee},\ and\ \citenamefont
  {Read}}]{Gardon}%
  \BibitemOpen
  \bibfield  {author} {\bibinfo {author} {\bibfnamefont {S.}~\bibnamefont
  {Gandon}}, \bibinfo {author} {\bibfnamefont {M.~J.}\ \bibnamefont
  {Mackinnon}}, \bibinfo {author} {\bibfnamefont {S.}~\bibnamefont {Nee}}, \
  and\ \bibinfo {author} {\bibfnamefont {A.~F.}\ \bibnamefont {Read}},\ }\href
  {\doibase 10.1038/414751a} {\bibfield  {journal} {\bibinfo  {journal}
  {Nature}\ }\textbf {\bibinfo {volume} {414}},\ \bibinfo {pages} {751}
  (\bibinfo {year} {2001})}\BibitemShut {NoStop}%
\bibitem [{\citenamefont {Kribs-Zaleta}\ and\ \citenamefont
  {Velasco-Hernández}(2000)}]{Zaleta}%
  \BibitemOpen
  \bibfield  {author} {\bibinfo {author} {\bibfnamefont {C.~M.}\ \bibnamefont
  {Kribs-Zaleta}}\ and\ \bibinfo {author} {\bibfnamefont {J.~X.}\ \bibnamefont
  {Velasco-Hernández}},\ }\href {\doibase
  https://doi.org/10.1016/S0025-5564(00)00003-1} {\bibfield  {journal}
  {\bibinfo  {journal} {Mathematical Biosciences}\ }\textbf {\bibinfo {volume}
  {164}},\ \bibinfo {pages} {183 } (\bibinfo {year} {2000})}\BibitemShut
  {NoStop}%
\bibitem [{\citenamefont {Peng}\ \emph {et~al.}(2013)\citenamefont {Peng},
  \citenamefont {Xu}, \citenamefont {Fu},\ and\ \citenamefont {Zhou}}]{Peng}%
  \BibitemOpen
  \bibfield  {author} {\bibinfo {author} {\bibfnamefont {X.-L.}\ \bibnamefont
  {Peng}}, \bibinfo {author} {\bibfnamefont {X.-J.}\ \bibnamefont {Xu}},
  \bibinfo {author} {\bibfnamefont {X.}~\bibnamefont {Fu}}, \ and\ \bibinfo
  {author} {\bibfnamefont {T.}~\bibnamefont {Zhou}},\ }\href {\doibase
  10.1103/PhysRevE.87.022813} {\bibfield  {journal} {\bibinfo  {journal} {Phys.
  Rev. E}\ }\textbf {\bibinfo {volume} {87}},\ \bibinfo {pages} {022813}
  (\bibinfo {year} {2013})}\BibitemShut {NoStop}%
\bibitem [{\citenamefont {Peng}\ \emph {et~al.}(2016)\citenamefont {Peng},
  \citenamefont {Xu}, \citenamefont {Small}, \citenamefont {Fu},\ and\
  \citenamefont {Jin}}]{Peng2}%
  \BibitemOpen
  \bibfield  {author} {\bibinfo {author} {\bibfnamefont {X.-L.}\ \bibnamefont
  {Peng}}, \bibinfo {author} {\bibfnamefont {X.-J.}\ \bibnamefont {Xu}},
  \bibinfo {author} {\bibfnamefont {M.}~\bibnamefont {Small}}, \bibinfo
  {author} {\bibfnamefont {X.}~\bibnamefont {Fu}}, \ and\ \bibinfo {author}
  {\bibfnamefont {Z.}~\bibnamefont {Jin}},\ }\href {\doibase
  10.1007/s00285-016-1007-3} {\bibfield  {journal} {\bibinfo  {journal}
  {Journal of Mathematical Biology}\ }\textbf {\bibinfo {volume} {73}},\
  \bibinfo {pages} {1561} (\bibinfo {year} {2016})}\BibitemShut {NoStop}%
\bibitem [{\citenamefont {Steinegger}\ \emph {et~al.}(2018)\citenamefont
  {Steinegger}, \citenamefont {Cardillo}, \citenamefont {Rios}, \citenamefont
  {G\'omez-Garde\~nes},\ and\ \citenamefont {Arenas}}]{Steinegger}%
  \BibitemOpen
  \bibfield  {author} {\bibinfo {author} {\bibfnamefont {B.}~\bibnamefont
  {Steinegger}}, \bibinfo {author} {\bibfnamefont {A.}~\bibnamefont
  {Cardillo}}, \bibinfo {author} {\bibfnamefont {P.~D.~L.}\ \bibnamefont
  {Rios}}, \bibinfo {author} {\bibfnamefont {J.}~\bibnamefont
  {G\'omez-Garde\~nes}}, \ and\ \bibinfo {author} {\bibfnamefont
  {A.}~\bibnamefont {Arenas}},\ }\href {\doibase 10.1103/PhysRevE.97.032308}
  {\bibfield  {journal} {\bibinfo  {journal} {Phys. Rev. E}\ }\textbf {\bibinfo
  {volume} {97}},\ \bibinfo {pages} {032308} (\bibinfo {year}
  {2018})}\BibitemShut {NoStop}%
\bibitem [{\citenamefont {Chen}\ and\ \citenamefont {Fu}(2019)}]{Chen}%
  \BibitemOpen
  \bibfield  {author} {\bibinfo {author} {\bibfnamefont {X.}~\bibnamefont
  {Chen}}\ and\ \bibinfo {author} {\bibfnamefont {F.}~\bibnamefont {Fu}},\
  }\href {\doibase 10.1098/rspb.2018.2406} {\bibfield  {journal} {\bibinfo
  {journal} {Proceedings of the Royal Society B: Biological Sciences}\ }\textbf
  {\bibinfo {volume} {286}},\ \bibinfo {pages} {20182406} (\bibinfo {year}
  {2019})},\ \Eprint
  {http://arxiv.org/abs/https://royalsocietypublishing.org/doi/pdf/10.1098/rspb.2018.2406}
  {https://royalsocietypublishing.org/doi/pdf/10.1098/rspb.2018.2406}
  \BibitemShut {NoStop}%
\bibitem [{\citenamefont {Wang}\ \emph {et~al.}(2019)\citenamefont {Wang},
  \citenamefont {Liu}, \citenamefont {Liang}, \citenamefont {Hu},\ and\
  \citenamefont {Zhou}}]{Wang2}%
  \BibitemOpen
  \bibfield  {author} {\bibinfo {author} {\bibfnamefont {W.}~\bibnamefont
  {Wang}}, \bibinfo {author} {\bibfnamefont {Q.-H.}\ \bibnamefont {Liu}},
  \bibinfo {author} {\bibfnamefont {J.}~\bibnamefont {Liang}}, \bibinfo
  {author} {\bibfnamefont {Y.}~\bibnamefont {Hu}}, \ and\ \bibinfo {author}
  {\bibfnamefont {T.}~\bibnamefont {Zhou}},\ }\href {\doibase
  https://doi.org/10.1016/j.physrep.2019.07.001} {\bibfield  {journal}
  {\bibinfo  {journal} {Physics Reports}\ }\textbf {\bibinfo {volume} {820}},\
  \bibinfo {pages} {1 } (\bibinfo {year} {2019})},\ \bibinfo {note}
  {coevolution spreading in complex networks}\BibitemShut {NoStop}%
\bibitem [{\citenamefont {Karrer}\ and\ \citenamefont {Newman}(2011)}]{Karrer}%
  \BibitemOpen
  \bibfield  {author} {\bibinfo {author} {\bibfnamefont {B.}~\bibnamefont
  {Karrer}}\ and\ \bibinfo {author} {\bibfnamefont {M.~E.~J.}\ \bibnamefont
  {Newman}},\ }\href {\doibase 10.1103/PhysRevE.84.036106} {\bibfield
  {journal} {\bibinfo  {journal} {Phys. Rev. E}\ }\textbf {\bibinfo {volume}
  {84}},\ \bibinfo {pages} {036106} (\bibinfo {year} {2011})}\BibitemShut
  {NoStop}%
\bibitem [{\citenamefont {Newman}(2005)}]{Newman2}%
  \BibitemOpen
  \bibfield  {author} {\bibinfo {author} {\bibfnamefont {M.~E.~J.}\
  \bibnamefont {Newman}},\ }\href {\doibase 10.1103/PhysRevLett.95.108701}
  {\bibfield  {journal} {\bibinfo  {journal} {Phys. Rev. Lett.}\ }\textbf
  {\bibinfo {volume} {95}},\ \bibinfo {pages} {108701} (\bibinfo {year}
  {2005})}\BibitemShut {NoStop}%
\bibitem [{\citenamefont {Cai}\ \emph {et~al.}(2015)\citenamefont {Cai},
  \citenamefont {Chen}, \citenamefont {Ghanbarnejad},\ and\ \citenamefont
  {Grassberger}}]{Cai}%
  \BibitemOpen
  \bibfield  {author} {\bibinfo {author} {\bibfnamefont {W.}~\bibnamefont
  {Cai}}, \bibinfo {author} {\bibfnamefont {L.}~\bibnamefont {Chen}}, \bibinfo
  {author} {\bibfnamefont {F.}~\bibnamefont {Ghanbarnejad}}, \ and\ \bibinfo
  {author} {\bibfnamefont {P.}~\bibnamefont {Grassberger}},\ }\href {\doibase
  10.1038/nphys3457} {\bibfield  {journal} {\bibinfo  {journal} {Nature
  physics}\ }\textbf {\bibinfo {volume} {11}},\ \bibinfo {pages} {936}
  (\bibinfo {year} {2015})}\BibitemShut {NoStop}%
\bibitem [{\citenamefont {Cui}\ \emph {et~al.}(2017)\citenamefont {Cui},
  \citenamefont {Colaiori},\ and\ \citenamefont {Castellano}}]{Cui}%
  \BibitemOpen
  \bibfield  {author} {\bibinfo {author} {\bibfnamefont {P.-B.}\ \bibnamefont
  {Cui}}, \bibinfo {author} {\bibfnamefont {F.}~\bibnamefont {Colaiori}}, \
  and\ \bibinfo {author} {\bibfnamefont {C.}~\bibnamefont {Castellano}},\
  }\href {\doibase 10.1103/PhysRevE.96.022301} {\bibfield  {journal} {\bibinfo
  {journal} {Phys. Rev. E}\ }\textbf {\bibinfo {volume} {96}},\ \bibinfo
  {pages} {022301} (\bibinfo {year} {2017})}\BibitemShut {NoStop}%
\bibitem [{\citenamefont {Min}\ and\ \citenamefont {Castellano}(2019)}]{Min}%
  \BibitemOpen
  \bibfield  {author} {\bibinfo {author} {\bibfnamefont {B.}~\bibnamefont
  {Min}}\ and\ \bibinfo {author} {\bibfnamefont {C.}~\bibnamefont
  {Castellano}},\ }\href@noop {} {\enquote {\bibinfo {title} {Message-passing
  theory for cooperative epidemics},}\ } (\bibinfo {year} {2019}),\ \Eprint
  {http://arxiv.org/abs/1912.01179} {arXiv:1912.01179 [physics.soc-ph]}
  \BibitemShut {NoStop}%
\bibitem [{\citenamefont {Funk}\ \emph {et~al.}(2009)\citenamefont {Funk},
  \citenamefont {Gilad}, \citenamefont {Watkins},\ and\ \citenamefont
  {Jansen}}]{Funk}%
  \BibitemOpen
  \bibfield  {author} {\bibinfo {author} {\bibfnamefont {S.}~\bibnamefont
  {Funk}}, \bibinfo {author} {\bibfnamefont {E.}~\bibnamefont {Gilad}},
  \bibinfo {author} {\bibfnamefont {C.}~\bibnamefont {Watkins}}, \ and\
  \bibinfo {author} {\bibfnamefont {V.~A.~A.}\ \bibnamefont {Jansen}},\ }\href
  {\doibase 10.1073/pnas.0810762106} {\bibfield  {journal} {\bibinfo  {journal}
  {Proceedings of the National Academy of Sciences}\ }\textbf {\bibinfo
  {volume} {106}},\ \bibinfo {pages} {6872} (\bibinfo {year} {2009})},\ \Eprint
  {http://arxiv.org/abs/https://www.pnas.org/content/106/16/6872.full.pdf}
  {https://www.pnas.org/content/106/16/6872.full.pdf} \BibitemShut {NoStop}%
\bibitem [{\citenamefont {Ruan}\ \emph {et~al.}(2012)\citenamefont {Ruan},
  \citenamefont {Tang},\ and\ \citenamefont {Liu}}]{Ruan}%
  \BibitemOpen
  \bibfield  {author} {\bibinfo {author} {\bibfnamefont {Z.}~\bibnamefont
  {Ruan}}, \bibinfo {author} {\bibfnamefont {M.}~\bibnamefont {Tang}}, \ and\
  \bibinfo {author} {\bibfnamefont {Z.}~\bibnamefont {Liu}},\ }\href {\doibase
  10.1103/PhysRevE.86.036117} {\bibfield  {journal} {\bibinfo  {journal} {Phys.
  Rev. E}\ }\textbf {\bibinfo {volume} {86}},\ \bibinfo {pages} {036117}
  (\bibinfo {year} {2012})}\BibitemShut {NoStop}%
\bibitem [{\citenamefont {Darabi~Sahneh}\ and\ \citenamefont
  {Scoglio}(2014)}]{Sahneh}%
  \BibitemOpen
  \bibfield  {author} {\bibinfo {author} {\bibfnamefont {F.}~\bibnamefont
  {Darabi~Sahneh}}\ and\ \bibinfo {author} {\bibfnamefont {C.}~\bibnamefont
  {Scoglio}},\ }\href {\doibase 10.1103/PhysRevE.89.062817} {\bibfield
  {journal} {\bibinfo  {journal} {Phys. Rev. E}\ }\textbf {\bibinfo {volume}
  {89}},\ \bibinfo {pages} {062817} (\bibinfo {year} {2014})}\BibitemShut
  {NoStop}%
\bibitem [{\citenamefont {Granell}\ \emph {et~al.}(2013)\citenamefont
  {Granell}, \citenamefont {G\'omez},\ and\ \citenamefont {Arenas}}]{Granell}%
  \BibitemOpen
  \bibfield  {author} {\bibinfo {author} {\bibfnamefont {C.}~\bibnamefont
  {Granell}}, \bibinfo {author} {\bibfnamefont {S.}~\bibnamefont {G\'omez}}, \
  and\ \bibinfo {author} {\bibfnamefont {A.}~\bibnamefont {Arenas}},\ }\href
  {\doibase 10.1103/PhysRevLett.111.128701} {\bibfield  {journal} {\bibinfo
  {journal} {Phys. Rev. Lett.}\ }\textbf {\bibinfo {volume} {111}},\ \bibinfo
  {pages} {128701} (\bibinfo {year} {2013})}\BibitemShut {NoStop}%
\bibitem [{\citenamefont {Sanz}\ \emph {et~al.}(2014)\citenamefont {Sanz},
  \citenamefont {Xia}, \citenamefont {Meloni},\ and\ \citenamefont
  {Moreno}}]{Sanz}%
  \BibitemOpen
  \bibfield  {author} {\bibinfo {author} {\bibfnamefont {J.}~\bibnamefont
  {Sanz}}, \bibinfo {author} {\bibfnamefont {C.-Y.}\ \bibnamefont {Xia}},
  \bibinfo {author} {\bibfnamefont {S.}~\bibnamefont {Meloni}}, \ and\ \bibinfo
  {author} {\bibfnamefont {Y.}~\bibnamefont {Moreno}},\ }\href {\doibase
  10.1103/PhysRevX.4.041005} {\bibfield  {journal} {\bibinfo  {journal} {Phys.
  Rev. X}\ }\textbf {\bibinfo {volume} {4}},\ \bibinfo {pages} {041005}
  (\bibinfo {year} {2014})}\BibitemShut {NoStop}%
\bibitem [{\citenamefont {Azimi-Tafreshi}(2016)}]{Azimi}%
  \BibitemOpen
  \bibfield  {author} {\bibinfo {author} {\bibfnamefont {N.}~\bibnamefont
  {Azimi-Tafreshi}},\ }\href {\doibase 10.1103/PhysRevE.93.042303} {\bibfield
  {journal} {\bibinfo  {journal} {Phys. Rev. E}\ }\textbf {\bibinfo {volume}
  {93}},\ \bibinfo {pages} {042303} (\bibinfo {year} {2016})}\BibitemShut
  {NoStop}%
\bibitem [{\citenamefont {Jo}\ \emph {et~al.}(2006)\citenamefont {Jo},
  \citenamefont {Baek},\ and\ \citenamefont {Moon}}]{Hang}%
  \BibitemOpen
  \bibfield  {author} {\bibinfo {author} {\bibfnamefont {H.-H.}\ \bibnamefont
  {Jo}}, \bibinfo {author} {\bibfnamefont {S.~K.}\ \bibnamefont {Baek}}, \ and\
  \bibinfo {author} {\bibfnamefont {H.-T.}\ \bibnamefont {Moon}},\ }\href
  {\doibase https://doi.org/10.1016/j.physa.2005.06.074} {\bibfield  {journal}
  {\bibinfo  {journal} {Physica A: Statistical Mechanics and its Applications}\
  }\textbf {\bibinfo {volume} {361}},\ \bibinfo {pages} {534 } (\bibinfo {year}
  {2006})}\BibitemShut {NoStop}%
\bibitem [{\citenamefont {Br\'odka}\ \emph {et~al.}(2020)\citenamefont
  {Br\'odka}, \citenamefont {Musial},\ and\ \citenamefont {Jankowsk}}]{Brodka}%
  \BibitemOpen
  \bibfield  {author} {\bibinfo {author} {\bibfnamefont {P.}~\bibnamefont
  {Br\'odka}}, \bibinfo {author} {\bibfnamefont {K.}~\bibnamefont {Musial}}, \
  and\ \bibinfo {author} {\bibfnamefont {J.}~\bibnamefont {Jankowsk}},\ }\href
  {\doibase 10.1109/ACCESS.2020.2965547} {\bibfield  {journal} {\bibinfo
  {journal} {IEEE Access}\ }\textbf {\bibinfo {volume} {8}},\ \bibinfo {pages}
  {10316} (\bibinfo {year} {2020})}\BibitemShut {NoStop}%
\bibitem [{\citenamefont {Yang}\ and\ \citenamefont {Li}(2016)}]{Yang2016}%
  \BibitemOpen
  \bibfield  {author} {\bibinfo {author} {\bibfnamefont {J.}~\bibnamefont
  {Yang}}\ and\ \bibinfo {author} {\bibfnamefont {C.-H.}\ \bibnamefont {Li}},\
  }\href {\doibase 10.1088/1751-8113/49/21/215601} {\bibfield  {journal}
  {\bibinfo  {journal} {Journal of Physics A: Mathematical and Theoretical}\
  }\textbf {\bibinfo {volume} {49}},\ \bibinfo {pages} {215601} (\bibinfo
  {year} {2016})}\BibitemShut {NoStop}%
\bibitem [{\citenamefont {Wu}\ \emph {et~al.}(2011)\citenamefont {Wu},
  \citenamefont {Fu},\ and\ \citenamefont {Yang}}]{Wu2011}%
  \BibitemOpen
  \bibfield  {author} {\bibinfo {author} {\bibfnamefont {Q.-C.}\ \bibnamefont
  {Wu}}, \bibinfo {author} {\bibfnamefont {X.-C.}\ \bibnamefont {Fu}}, \ and\
  \bibinfo {author} {\bibfnamefont {M.}~\bibnamefont {Yang}},\ }\href {\doibase
  10.1088/1674-1056/20/4/046401} {\bibfield  {journal} {\bibinfo  {journal}
  {Chinese Physics B}\ }\textbf {\bibinfo {volume} {20}},\ \bibinfo {pages}
  {046401} (\bibinfo {year} {2011})}\BibitemShut {NoStop}%
\bibitem [{\citenamefont {Wu}\ \emph {et~al.}(2013)\citenamefont {Wu},
  \citenamefont {Small},\ and\ \citenamefont {Liu}}]{Wu2013}%
  \BibitemOpen
  \bibfield  {author} {\bibinfo {author} {\bibfnamefont {Q.}~\bibnamefont
  {Wu}}, \bibinfo {author} {\bibfnamefont {M.}~\bibnamefont {Small}}, \ and\
  \bibinfo {author} {\bibfnamefont {H.}~\bibnamefont {Liu}},\ }\href {\doibase
  10.1007/s00332-012-9146-1} {\bibfield  {journal} {\bibinfo  {journal}
  {Journal of Nonlinear Science}\ }\textbf {\bibinfo {volume} {23}},\ \bibinfo
  {pages} {113} (\bibinfo {year} {2013})}\BibitemShut {NoStop}%
\bibitem [{\citenamefont {Ahn}\ \emph {et~al.}(2006)\citenamefont {Ahn},
  \citenamefont {Jeong}, \citenamefont {Masuda},\ and\ \citenamefont
  {Noh}}]{Ahn2006}%
  \BibitemOpen
  \bibfield  {author} {\bibinfo {author} {\bibfnamefont {Y.-Y.}\ \bibnamefont
  {Ahn}}, \bibinfo {author} {\bibfnamefont {H.}~\bibnamefont {Jeong}}, \bibinfo
  {author} {\bibfnamefont {N.}~\bibnamefont {Masuda}}, \ and\ \bibinfo {author}
  {\bibfnamefont {J.~D.}\ \bibnamefont {Noh}},\ }\href {\doibase
  10.1103/PhysRevE.74.066113} {\bibfield  {journal} {\bibinfo  {journal} {Phys.
  Rev. E}\ }\textbf {\bibinfo {volume} {74}},\ \bibinfo {pages} {066113}
  (\bibinfo {year} {2006})}\BibitemShut {NoStop}%
\bibitem [{\citenamefont {Szolnoki}\ \emph {et~al.}(2014)\citenamefont
  {Szolnoki}, \citenamefont {Mobilia}, \citenamefont {Jiang}, \citenamefont
  {Szczesny}, \citenamefont {Rucklidge},\ and\ \citenamefont {Perc}}]{RSP}%
  \BibitemOpen
  \bibfield  {author} {\bibinfo {author} {\bibfnamefont {A.}~\bibnamefont
  {Szolnoki}}, \bibinfo {author} {\bibfnamefont {M.}~\bibnamefont {Mobilia}},
  \bibinfo {author} {\bibfnamefont {L.-L.}\ \bibnamefont {Jiang}}, \bibinfo
  {author} {\bibfnamefont {B.}~\bibnamefont {Szczesny}}, \bibinfo {author}
  {\bibfnamefont {A.~M.}\ \bibnamefont {Rucklidge}}, \ and\ \bibinfo {author}
  {\bibfnamefont {M.}~\bibnamefont {Perc}},\ }\href {\doibase
  10.1098/rsif.2014.0735} {\bibfield  {journal} {\bibinfo  {journal} {J. R.
  Soc. Interface}\ }\textbf {\bibinfo {volume} {11}},\ \bibinfo {pages}
  {20140735} (\bibinfo {year} {2014})}\BibitemShut {NoStop}%
\end{thebibliography}%


\begin{thebibliography}{99}


\bibitem{Newman}
M. E. J. Newman, {\em Spread of epidemic disease on networks}, Phys. Rev. E. {\bf 66}, 016128 (2002).


\bibitem{castellano}
C. Castellano, S. Fortunato, and V. Loreto, {\em Statistical physics of social dynamics}, Reviews of Modern Physics {\bf 81}, 591 (2009).


\bibitem{vaccine-review}
Z. Wang, C. T. Bauch, S. Bhattacharyya, A. d\'Onofrio, P. Manfredi, M. Perc, N. Perra, M. Salath\'e, D. Zhao,  {\em Statistical physics of vaccination}, Physics Reports {\bf 664} 1-113 (2016).

\bibitem{Pastor}
R. Pastor-Satorras and A. Vespignani, {\em Immunization of complex networks}, Phys. Rev. E {\bf 65}, 036104 (2002).

\bibitem{Inverse}
M. Schneider, T. Mihalijev, and H. J. Herrmann, {\em Inverse targeting —An effective immunization strategy},  Europhys. Lett. {\bf 98}, 46002 (2012).

\bibitem{Wang}
Z. Wang, Y. Moreno,S. Boccaletti, M. Perc, {\em Vaccination and epidemics in networked populations—An introduction}, Chaos, Solitons and Fractals {\bf 103} 177 (2017).

\bibitem{target}
C. M. Schneider,T. Mihaljev, S. Havlin, H. J. Herrmann, {\em Suppressing epidemics with a limited amount of immunization units}, Phys. Rev. E. {\bf 84} 061911 (2011).


\bibitem{Acquaintance}
R. Cohen, S. Havlin, and D. ben-Avraham, {\em Efficient Immunization Strategies for Computer Networks and Populations}, Phys. Rev. Lett. {\bf 91}, 247901 (2003).

\bibitem{Gardon}
S. Gandon, M. J. Mackinnon, S. Nee, A. F. Read, {\em Imperfect vaccines and the evolution of pathogen
virulence}, Nature {\bf 414}, 751 (2001).


\bibitem{Zaleta}
C. M. Kribs-Zaleta and J. X. Velasco-Hernandez, {\em A simple vaccination model with multiple endemic states}, Math. Biosci. {\bf 164}, 183 (2000).


\bibitem{Peng}
 X.-L. Peng,  X.-J. Xu,  X. Fu, and T. Zhou, {\em Vaccination intervention on epidemic dynamics in networks}, Phys. Rev. E. {\bf 87} 022813 (2013).


\bibitem{Peng2}
 X.-L. Peng , X.-J. Xu, M. Small, X. Fu, Z. Jin, {\em Prevention of infectious diseases by public vaccination and individual protection}, Journal of Mathematical Biology {\bf 73}, 1561 (2016).


\bibitem{Karrer}
B. Karrer and M. E. J. Newman, {\em Competing Epidemics on Complex Networks}, Phys. Rev. E. {\bf 84}, 036106 (2011).

\bibitem{Newman2}
M. E. J. Newman, {\em Threshold Effects for Two Pathogens Spreading on a Network}, Phys. Rev. Lett. {\bf 95}, 108701 (2005).

\bibitem{Cai}
W. Cai, L. Chen, F.Ghanbarnejad, and P. Grassberger, {\em Avalanche outbreaks emerging in cooperative contagions}, Nature Phys. {\bf 11}, 936 (2015).


\bibitem{Funk}
S. Funk, E. Gilad, C. Watkins, and V. A. A Jansen, {\em The spread of awareness and its impact onepidemic outbreaks}, Proceedings of the National Academy of Sciences {\bf 106}, 16, 6872 (2009).


\bibitem{Ruan}
Z. Ruan, M. Tang, and Z. Liu, {\em Epidemic spreading with information-driven vaccination}, Phys. Rev. E. {\bf 86}, {\bf 3}  036117 (2012).


\bibitem{Sahneh}
F. D. Sahneh and C. Scoglio, {\em Competitive Epidemic Spreading Over Arbitrary Multilayer Networks}, Phys. Rev. E. {\bf 89}, 062817 (2014).

\bibitem{Granell}
C. Granell, S. G\'omez, and A. Arenas, {\em Dynamical Interplay between Awareness and Epidemic Spreading in Multiplex Networks}, Phys. Rev. Lett. {\bf 111}, 128701 (2013).

\bibitem{Sanz}
J. Sanz, Ch-Y Xia, S. Meloni, and Y. Moreno, {\em Dynamics of Interacting Diseases}, Phys, Rev. X. {\bf 4}, 041005 (2014).

\bibitem{Azimi}
N. Azimi-Tafreshi, {\em Cooperative epidemics on multiplex networks}, Phys. Rev. E. {\bf 93}, 042303 (2016).

\bibitem{Hang}
H-H. Jo, S. K. Baek, and H-T. Moon, {\em Immunization dynamics on a two-layer network model}, Physica A: Statistical Mechanics and its Applications 361 {\bf 2}, 534 (2006)


\bibitem{Brodka}
P. Br\'odka, K. Musial, J. Jankowski, {\em Interacting spreading processes in multilayer networks}, arXiv:1903.05932 (2019).



\end{thebibliography}

\end{document}